\documentclass[twocolumn]{emulateapj}

\usepackage{amsmath}

\shortauthors{
Wong, Sarazin, \& Ji}

\shorttitle{
NEI Signatures in Cluster Outerskirts
}
\slugcomment{Astrophysical Journal, accepted}

\begin{document}
\title{
X-ray Signatures of Non-Equilibrium Ionization Effects in Galaxy Cluster 
Accretion Shock Regions
}

\author{Ka-Wah Wong\altaffilmark{1,2},
Craig L. Sarazin\altaffilmark{2},
and
Li Ji\altaffilmark{3}
}

\altaffiltext{1}{Department of Physics and Astronomy, University of 
Alabama, Box 870324, Tuscaloosa, AL 35487, USA
}
\altaffiltext{2}{Department of Astronomy, University of Virginia,
P. O. Box 400325, Charlottesville, VA 22904-4325, USA}
\altaffiltext{3}{MIT Kavli Institute for Astrophysics and Space 
Research, Cambridge, MA 02139, USA
}

\email{kwong@ua.edu, sarazin@virginia.edu, ji@space.mit.edu}

\begin{abstract}

The densities in the outer regions of clusters of 
galaxies are very low, and the collisional timescales are very long.
As a result, heavy elements will be 
under-ionized after they have passed through the accretion shock. 
We have 
studied systematically the effects of non-equilibrium ionization for 
relaxed clusters in the $\Lambda$CDM cosmology using one-dimensional 
hydrodynamic simulations.  We found that non-equilibrium ionization 
effects do not depend on cluster mass but depend strongly on redshift 
which can be understood by self-similar scaling arguments.  The effects 
are 
stronger for clusters at lower redshifts.  We present X-ray signatures 
such as surface brightness profiles and emission lines in detail for a 
massive cluster at low redshift.  In general, soft emission
(0.3--1.0~keV) is enhanced significantly by under-ionization, and 
the enhancement can be nearly an order of magnitude near the shock 
radius.  The most prominent non-equilibrium ionization signature we found 
is the \ion{O}{7} and \ion{O}{8} line ratio.  
The ratios for 
non-equilibrium ionization and collisional ionization equilibrium models 
are different by more than an order of magnitude at radii beyond 
half of the shock radius.  These non-equilibrium ionization 
signatures are equally strong for models with different non-adiabatic 
shock electron heating efficiencies.  We have also calculated the
detectability of the \ion{O}{7} and \ion{O}{8} lines with the future
{\it International X-ray Observatory} ({\it IXO}).  Depending on the line 
ratio measured, we conclude that 
an exposure 
of $\sim$130--380~ksec on a moderate-redshift, massive regular cluster
with the X-ray Microcalorimeter Spectrometer (XMS) on 
the {\it IXO} will be sufficient to provide a strong test for the 
non-equilibrium ionization model.

\end{abstract}


\section{Introduction}
\label{ion_sec:intro}

Clusters of galaxies are very sensitive probes to cosmological parameters 
\citep[e.g.,][]{ARS08,Vik+09}.  Systematic uncertainties in precision 
cosmology 
using galaxy clusters can be minimized by restricting the sample of 
clusters to the highest degree of dynamical relaxation, and hence 
studying relaxed clusters is particularly important.  In addition, the 
outer envelopes of clusters have been thought to be less subjected to 
complicated physics such as active galactic nucleus feedback, and hence 
these outer regions may provide better cosmological probes.  

Clusters are believed to be formed by the continual merging and 
accretion of material
from the surrounding large-scale structure. 
Based on high resolution $N$-body simulations, it 
has been found that cluster growth can be divided into an early fast 
accretion phase dominated by major mergers, and a late slow phase 
dominated by smooth accretion of background materials and many minor 
mergers \citep{WBP+02,ZJM+09}.  Accretion shocks (or virial shocks) are 
unambiguous predictions of cosmological hydrodynamic simulations.  For 
the most relaxed clusters with roughly spherical morphology in the outer 
regions, these simulations show that large amounts of material are accreted 
through filamentary structures in some particular directions, and more 
spherically symmetrically in other directions.  

Most of the simulations assume the intracluster medium (ICM) is in 
collisional equilibrium.  However, because of the very low density in the 
cluster outer regions ($\ga R_{200}$\footnote{$R_{\Delta}$ is the 
radius within which the mean total mass density of the cluster is 
$\Delta$ times the critical density of the universe.  The virial radius 
$R_{\rm vir}$ is defined as a radius within which the cluster is 
virialized.  For the Einstein-de Sitter universe, $ R_{\rm vir} \approx 
R_{178}$, while for the standard $\Lambda$CDM Universe, $ R_{\rm vir}  
\approx R_{95}$.}), the 
Coulomb collisional and collisional ionization timescales are comparable to 
the age of the cluster.  It has been pointed out that electrons and ions 
there may be in non-equipartition \citep{FL97,Tak99,Tak00,RN09,WS09}, and 
also the 
ionization state may not be in collisional ionization equilibrium 
\citep{YYS+03,CF06,YS06}.  
If these 
non-equilibrium processes are not properly taken into account, the 
measured properties may be biased in these regions.

Studying the outer regions of relaxed clusters not only is valuable in 
understanding the accretion physics and to test the assumptions 
concerning plasma 
physics near the shock regions, but it is also very important to test 
structure formation theory and to constrain the systematic uncertainties 
in clusters to be used as precision cosmological probes.  Before the 
launch of the {\it Suzaku} X-ray observatory, physical properties such as 
temperature have never been constrained with confidence beyond roughly 
one-half of the shock radius.  Most of our understanding of these regions 
is still based on theoretical models and hydrodynamic simulations.
Recently, observations by {\it Suzaku} have constrained
temperatures up to about half of the shock radius for a few clusters for 
the first time to better than a factor of $\sim 2$ 
\citep{GFS+09,Rei+09, Bas+10,Hos+10}.  While the uncertainties are still 
large, there is evidence that the electron pressure in cluster 
outer regions may be lower than that predicted by numerical simulations 
assuming collisional equilibrium.  Observations of secondary cosmic 
microwave background anisotropies with the South Pole Telescope (SPT) and 
the 
{\it Wilkinson Microwave Anisotropy Probe} ({\it WMAP}) 7 year data also 
support these results \citep{Kom+10,Lue+10}.  These observational 
signatures are consistent with electrons and ions in non-equipartition, 
although it is also possible that the hydrodynamic simulations may simply 
overestimate the gas pressure.  Another possibility is that heat 
conduction outside the cluster may be reducing the gas pressure 
\citep{Loe02}.  Recently, \citet{WSW10} have shown that cosmological 
parameters will be biased if non-equilibrium effects
(in particular, non-equipartition) are not properly taken into account.

More observations of the outer regions of clusters are being done or analyzed.
In the future, the proposed {\it International X-ray Observatory} 
($IXO$)\footnote{http://ixo.gsfc.nasa.gov/} will have the sensitivity to 
constrain cluster properties out to the shock radius.
Thus, a detailed study
of the physics in the outer regions of clusters would be useful.  In 
particular, {\it IXO} 
will have sufficient spectral resolution to resolve many important X-ray
lines, and hence the ionization state of the plasma  can be determined.
 
In \citet{WS09}, we have studied in detail the X-ray signature of 
non-equipartition effects in cluster accretion shock regions.  In this 
paper, we extend our study to include non-equilibrium ionization in our 
calculations.  Non-equilibrium ionization calculations have been 
considered in a number of cosmological simulations to study the very low 
density warm-hot intergalactic medium (WHIM) surrounding galaxy clusters 
\citep{YYS+03,CF06,YS06}.  
At galaxy cluster scales, similar 
non-equilibrium ionization calculations
have focused on merging clusters \citep{AY08,AY10}.  These studies generally 
agree that in the low density ICM and the WHIM, there are significant 
deviations from ionization equilibrium, and that the effects on the X-ray 
emission lines are strong.  In this work, we focus on the X-ray 
signatures of non-equilibrium ionization in the accretion shock regions 
of relaxed clusters.
We also discuss the detectability of non-equilibrium 
ionization effects with the future {\it IXO}.

The paper is organized as follows.  Section~\ref{ion_sec:method} 
describes in detail the physical models and techniques to calculate 
non-equilibrium ion fractions and X-ray observables, which includes the
hydrodynamic models we used (Section~\ref{ion_sec:hydro}), 
and the ionization and 
spectral calculational techniques 
(Sections~\ref{ion_sec:IonFracCal}--\ref{ion_sec:XrayCal}).
The overall dependence the non-equilibrium ionization effects on the 
mass and redshift of the cluster
is presented in Section~\ref{ion_sec:b_vs_mass_z}.
Calculated non-equilibrium 
ionization signatures are presented in Section~\ref{ion_sec:signatures}, 
which includes
descriptions of particular models we used to present the results 
(Section~\ref{ion_sec:models}), 
X-ray spectra (Section~\ref{ion_sec:spectra}), surface 
brightness profiles (Section~\ref{ion_sec:sb}), and the ratio of 
intensities of \ion{O}{7} and 
\ion{O}{8} lines (Section~\ref{ion_sec:line}).  We discuss the 
detectability of \ion{O}{7} and \ion{O}{8} lines and
non-equilibrium ionization diagnostics with {\it IXO} in 
Section~\ref{ion_sec:detect}.  Section~\ref{ion_sec:conclusion} gives the 
discussion and conclusions.  Throughout the paper, we assume a Hubble 
constant of $H_0 = 71.9~h_{71.9}$~km~s$^{-1}$~Mpc$^{-1}$ with $h_{71.9}=1$, 
a total matter density parameter of $\Omega_{M,0} = 0.258$, a dark 
energy density parameter of $\Omega_{\Lambda} = 0.742$, and a cluster 
gas fraction 
of $f_{\rm gas}=\Omega_b/\Omega_M=0.17$, where $\Omega_b$ is the baryon 
density parameter for our cluster models in the standard $\Lambda$CDM 
cosmology\footnote{\scriptsize
http://lambda.gsfc.nasa.gov/product/map/dr3/parameters\_summary.cfm}.
The clusters have a hydrogen mass fraction $X=76\%$ for the ICM.

\section{Physical Models and Calculational Techniques}
\label{ion_sec:method}

\subsection{Hydrodynamic Models and Electron Temperature Structures}
\label{ion_sec:hydro}

We calculate the X-ray emission spectrum from the outer regions of 
cluster of galaxies using 
one-dimensional hydrodynamic simulations we have developed \citep{WS09}.  
Radiative cooling is negligible in cluster outer regions so that it will 
not affect the dynamics of the plasma.  This non-radiative condition 
allows us to calculate the hydrodynamics and then the ionization 
structure and X-ray emission separately.
The hydrodynamic models simulate the accretion of background material 
from the surrounding regions onto clusters through accretion shocks in 
the $\Lambda$CDM cosmology.  The calculated hydrodynamic variables (e.g., 
density and temperature profiles) in the cluster outer regions are 
consistent with those calculated by three-dimensional simulations, and a 
detailed discussion on our hydrodynamic simulations can be found in 
\citet{WS09}.

The hydrodynamic simulations were done using Eulerian coordinates.  In 
calculating the time-dependent ionization for the hydrodynamic models,
we need to follow the Lagrangian history of each fluid 
element.  In one-dimensional spherical symmetric systems, it is 
convenient to transform from the Eulerian coordinates to the Lagrangian 
coordinates by using the interior gas mass as the independent variable
\begin{equation}
\label{ion_eq:coor}
m_g(r) = 4\pi \int_0^r \rho_g r^2 dr \, ,
\end{equation} 
where $r$ is the radius, and $\rho_g$ is the gas density. 
To do this, we first determined the values of $m_g(r)$ for each of the 
grid zones in the 
hydrodynamical simulations in \citet{WS09} at the final redshift of $z = 0$.
The values of the density and temperature were also determined for this 
final time step for each gas element.
Then, for each earlier time step, the values of gas density and 
temperature
within the shock radius
were determined by linear interpolation
between the nearest two values of the interior gas mass on the grid at 
that time.
When the values of $m_g(r)$ fall between the discontinuous values of the 
shocked and preshocked elements, we assume the material to be 
preshock.
This slightly underestimates the ionization timescale parameter
defined below, but this effect is small as the
grid zones in the simulations are closely spaced.

The collisional ionization and recombination rates and the excitation of 
X-ray emission depend on the electron temperature ($T_e$).  Because
the accretion shock may primarily heat ions instead of electrons due to the 
mass difference and the long Coulomb collisional timescale between 
electrons and ions, the electron temperature may not be the same as the 
ion temperature near the accretion shock regions \citep{FL97, WS09}. 
The degree of non-equipartition depends on the non-adiabatic shock 
electron heating efficiency, which is defined as $\beta \equiv \Delta 
T_{e,{\rm non{\text -}ad}} / \Delta {\bar T}_{\rm non{\text -ad}}$ in 
\citet{WS09}, 
where $\Delta T_{e,{\rm non{\text -}ad}}$ and $\Delta {\bar T}_{\rm 
non{\text -}ad}$ are the changes in electron 
temperature and average thermodynamic temperature due to non-adiabatic 
heating at the shock, respectively. 
While there are no observational constraints on 
$\beta$ in galaxy cluster accretion shocks, observations in supernova 
remnants suggest that $\beta \ll 1$ for shocks with Mach numbers 
similar to those in galaxy cluster accretion shocks \citep{GLR07}.  
\citet{WS09} have calculated electron temperature profiles for models 
with a very low electron heating efficiency ($\beta = 1/1800$), an 
intermediate electron heating efficiency ($\beta = 0.5$), and an 
equipartition model ($\beta = 1$).  The electron temperature profiles of 
these models with different values of $\beta$ are used to calculate the 
ionization fractions and X-ray emission in this paper.

\begin{figure}[h]
\includegraphics[angle=270,width=12cm]{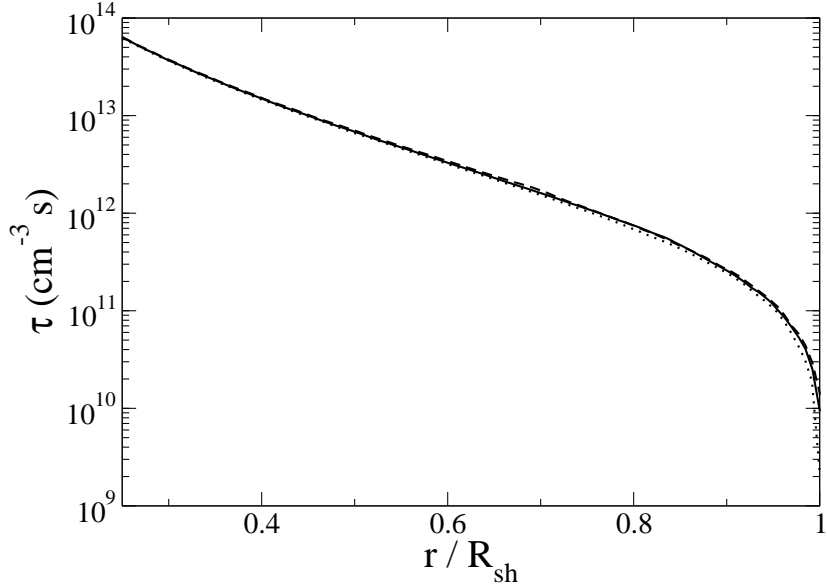}
\caption{
Ionization timescale parameter $\tau$ (Equation~\ref{ion_eq:iontime}) versus 
radius $r$ scaled to the cluster shock radius $R_{\rm sh}$ for cluster 
models with total accreted masses of $M_{\rm 
sh} = 0.77$ (dashed line), 1.53 (solid line), and 3.06 (dotted line) 
$\times$ $10^{15}~M_{\odot}$ at a redshift of $z=0$.
The shock radii for the three clusters from small to high mass are 
$R_{\rm sh} = 3.31, 4.22$, and 5.41~Mpc, respectively.
All three lines lie almost on top of one another.
}
\label{ion_fig:timescale}
\end{figure}

The ionization timescale parameter of each fluid element is defined as
\begin{equation}
\label{ion_eq:iontime}
\tau = \int_{t_s}^{t_0} n_e \, dt \,,
\end{equation}
where $n_e$ is the electron number density, $t_s$ is the time when the fluid 
element was shocked, and $t_0$ is the time at the observed redshift.
The ionization timescale parameters for cluster 
models with accreted masses of $M_{\rm sh} = 0.77, 1.53$, and $3.06 
\times 10^{15}~M_{\odot}$ are shown in Figure~\ref{ion_fig:timescale}.  
Most ions of astrophysical interests will not achieve ionization 
equilibrium for the ionization timescale parameters $\la 10^{12}$~cm$^{-3}$~s 
\citep{SH10}.

\subsection{Ionization Calculations}
\label{ion_sec:IonFracCal}

In order to calculate the X-ray emission spectrum for non-equilibrium 
ionization plasma, the ionization state of each fluid element has to be 
calculated.  We assume H and He are fully ionized. 

In this paper, we consider only collisional ionization processes, and 
ignore photoionization.
To justify this, we compare 
the photoionization and collisional ionization rate for the ions we are 
interested in.  The photoionization rate of an ion $i$ is given by
\begin{equation}
\label{ion_eq:PIrate}
R_{\rm PI} = \int_0^{\infty} \frac{\sigma_i(\nu) F(\nu)}{h \nu} d\nu \,,
\end{equation}
where $\sigma_i(\nu)$ is the photoionization cross section of ion $i$ at 
frequency $\nu$, and $F(\nu)$ is the ionizing flux.  Near the cluster 
accretion shock, the dominant photoionization source is the UV 
background. 
We approximate the UV ionizing flux at $z=0$ as a power law
$F( \nu ) = F_0 (\nu/\nu_0)^{-\Gamma}$ with $F_0 = 10^{22}$ 
erg~cm$^{-2}$~s$^{-1}$~Hz, $\nu_0 = 10^{15.5}$~Hz, and $\Gamma = 1.3$ 
\citep{HM96}.  For example, for the \ion{O}{7} and \ion{O}{8} ions 
we are most interested in throughout this paper, adopting the 
photoionization cross sections given by \citet{VFK+96} gives $R_{\rm 
PI}($\ion{O}{7}$) = 5.2 \times 10^{-18}$~s$^{-1}$ and $R_{\rm 
PI}($\ion{O}{8}$) = 1.7 \times 10^{-18}$~s$^{-1}$.  The slowest 
collisional 
ionization rate of the oxygen ions is of the order of $10^{-16} 
(n_e/10^{-5}~{\rm cm}^{-3}$)~s$^{-1}$ for typical
temperatures ($> 1$~keV) and densities in the outer regions of clusters 
\citep{SH10}.
This is nearly two orders 
of magnitude higher than the photoionization rates of \ion{O}{7} and
\ion{O}{8} ions.  Even at $z=3$ when the UV background is roughly 80 
times stronger \citep{HM96}, the densities in the outer regions of 
clusters will be higher by a factor of $\sim (1+z)^3 = 64$, and hence 
collisional ionization rates will increase by a similar factor.  
Collisional ionization still dominates over photoionization.  For heavier 
elements, the photoionization rates will be even smaller, and the 
collisional ionization rates for ions up to those of Ni are all higher 
than $\sim 10^{-17} (n_e/10^{-5}~{\rm cm}^{-3}$)~s$^{-1}$ \citep{SH10}.  
Therefore, we assume that photoionization is not important in our 
calculations.  

For ionization dominated by collisional processes at low densities, the 
ionization states 
for each of the ions of a given element $X$ are governed by
\begin{eqnarray}
\label{ion_eq:ionization}
\frac{df_i}{dt} & = & n_e \{ C_{i-1}(T_e) f_{i-1} + \alpha_i(T_e) f_{i+1} 
\nonumber \\
& &
\hskip +10mm
- [ C_i(T_e) + \alpha_{i-1}(T_e) ] f_i  \} \,,
\end{eqnarray}
where $f_i \equiv n( X^{+i-1} )/n(X)$  is the ionization fraction of the 
ion $i$ with charge $+i-1$, $n( X^{+i-1})$ is the number density of that 
ion, $n(X)$ is the total number density of the element $X$, and 
$C_{i}(T_e)$ and $\alpha_i(T_e)$ are the coefficients of collisional 
ionization out of and 
recombination into the ion $i$, respectively.
In solving Equation~(\ref{ion_eq:ionization}), we use an eigenfunction 
technique which is based on the algorithm developed by \citet{HH85}, and 
the method is described in detail in Appendix A of  \citet{BSB94}. 
The eleven heavy elements C, N, O, Ne, Mg, Si, S, Ar, Ca, Fe, and Ni are 
included in our calculations.  The eigenvalues and eigenvectors, which 
are related to the ionization and recombination rates, used to solve 
Equation~(\ref{ion_eq:ionization}) are taken from the latest version of 
the {\it nei} version 2 model in 
XSPEC\footnote{http://heasarc.nasa.gov/xanadu/xspec/} (version 12.6.0).  
The atomic physics used to calculated the eigenvalues and eigenvectors in 
XSPEC are discussed in \citet{BLR01}, and the ionization fractions in the 
latest version of XSPEC are calculated using the updated dielectronic 
recombination rates from \citet{MMC+98}.  
All of the eleven heavy elements are assumed to be neutral initially when 
solving Equation~(\ref{ion_eq:ionization}), which is a good approximation 
as long as the ionization states of the preshock plasma are much lower 
than that of the postshock plasma \citep[e.g.,][]{BSB94,BLR01,JWK06}.
We have also tested this by assuming all the ionization states in the 
preshock regions are in collisional ionization equilibrium initially at 
different temperatures of $10^4, 10^5$ and $10^6$~K.  We confirmed that 
the final ionization fractions for \ion{O}{7} and \ion{O}{8} are 
essentially the same.

\subsection{X-ray Emission Calculations}
\label{ion_sec:XrayCal}

Once the ionization states are calculated by solving 
Equation~(\ref{ion_eq:ionization}), the X-ray emission spectrum could be 
calculated by using available plasma emission codes such as the 
Raymond-Smith code \citep{RS77}, the SPEX code \citep{KMN96}, and the 
APEC code \citep{SBL+01}. 
We chose to use a version of the APEC code as implemented in the
{\it nei} version 2 model in XSPEC to 
calculate the X-ray emission spectrum.  The routine uses the 
Astrophysical Plasma Emission 
Database\footnote{http://cxc.harvard.edu/atomdb/} (APED) to calculate the 
resulting spectrum. 
Thus, the atomic data we used to calculate the ionization 
fractions and spectrum are mutually consistent, both coming from the 
same {\it nei} model version 2.0 in XSPEC.
The publicly available line list {\it APEC\_nei\_v11} was used 
throughout the paper.  The inner-shell processes are missing in this 
{\it nei} version 2.  We have compared our results to an updated line 
list which includes the inner-shell processes but is not yet publicly released 
(K. Borkowski, private communications), and we found that the difference 
is less than 10\% which is smaller than the $\sim 30\%$ uncertainties in 
the atomic physics of the X-ray plasma code.

We assume the abundance to be a typical value for galaxy clusters, which 
is 0.3 of solar for all models, and use the 
solar abundance tables of
\citet{AG89}.

We calculated the X-ray emissivity in photons per unit time per 
unit volume per unit energy, $\epsilon_E$, for each fluid element in our 
hydrodynamic models at redshift zero,
which can be expressed in terms of an emissivity function, $\Lambda_E$, 
by \citep{Sar86}
\begin{equation}
\label{ion_eq:emiss}
\epsilon_E = \Lambda_E \, n_e n_p \,,
\end{equation}
where $n_p$ is the proton density.
The emissivity function $\Lambda_E$ depends on the ionization fractions 
and the electron temperature, but is independent of the gas density.
The projected spectrum in the rest frame is given by
integrating the X-ray emissivity along the line of sight \citep{Sar86}
\begin{equation}
\label{ion_eq:spectrum}
I_E = \int \epsilon_E \, dl \,,
\end{equation}
where $l$ is the distance along the line of sight. 
The broadband rest frame surface brightness in energy per unit time per 
unit area is then given by
\begin{equation}
\label{ion_eq:sb}
S_E = \int I_E \, E \, dE \,,
\end{equation}
where the integral is across the energy band of interest.

\section{Dependence of Non-Equilibrium Ionization on Cluster Mass and Redshift}
\label{ion_sec:b_vs_mass_z}

We consider how the degree on non-equilibrium ionization of a cluster
depends on its mass and redshift.
We first consider how the cluster parameters which affect the ionization depend
on mass and redshift.
In order to compare equivalent locations in the different clusters, we estimate the
ionization parameters at a given over-density radius $R_\Delta$.
We consider a simple self-similar scaling argument for galaxy clusters. 

By definition, the average total density within the radius $R_\Delta$ is
\begin{equation}
\label{ion_eq:rhodelta}
\rho_\Delta
\equiv \Delta \, \rho_{\rm crit} (z) 
= \frac{3 \Delta}{8 \pi G} H^2 (z) 
= \frac{3 \Delta H_0^2}{8 \pi G} E^2 (z) \, ,
\end{equation}
where $\rho_{\rm crit} (z)$ is the critical density at redshift $z$, and
$ H(z)$ is the Hubble constant at redshift $z$.
The quantity $E(z) = H(z) / H_0 $, so that
\begin{eqnarray}
\label{ion_eq:Ez}
E^2 (z)
& = &
\left[
\Omega_M (1 + z)^3
+ \Omega_R (1+z)^4
+ \Omega_\Lambda
\right. \nonumber \\
&  &
\left.
+ \left( 1 - \Omega_M - \Omega_R - \Omega_\Lambda \right) (1+z)^2
\right] \nonumber \\
& \approx &
\Omega_M (1 + z)^3 + 1 - \Omega_M
\, .
\end{eqnarray}
The final expression in Equation~(\ref{ion_eq:Ez}) follows from the assumption that the universe is flat
($\Omega_M + \Omega_R + \Omega_\Lambda = 1$)
and the fact that the radiation density parameter is small ($\Omega_R \ll 1$) in the present-day universe.
The gas density at $R_\Delta$ is roughly
\begin{equation}
\label{ion_eq:rhogas}
\rho_{\rm gas} ( R_\Delta ) \approx f_{\rm gas} \rho_\Delta \propto  \Delta \, f_{\rm gas}  E^2 (z)
\, .
\end{equation}
The inflow timescale is roughly
\begin{equation}
\label{ion_eq:tflow}
t
\approx
\left( G \rho_\Delta \right)^{-1/2}
\propto  t_H (0) \Delta^{-1/2} [E(z)]^{-1}
\, ,
\end{equation}
where $t_H (0) \equiv 1 / H_0$ is the Hubble time at $z = 0$.
Thus, the ionization timescale parameter $\tau$ varies as
\begin{equation}
\label{ion_eq:tauscale}
\tau \propto \Delta^{1/2} \, f_{\rm gas}  E(z)
\, .
\end{equation}
Thus, the value of $\tau$ at a fixed characteristic radius should be 
nearly independent of the cluster mass
but will increase significantly with redshift.
Figure~\ref{ion_fig:timescale} shows that the variation of $\tau$ with 
radius is fairly self-similar for clusters with differing masses.
When these curves are scaled to a fixed cluster characteristic radius, 
they are very nearly identical.

The ionization state of the gas depends on $\tau$ and the collisional ionization and recombination rates.
For an under-ionized plasma, the collisional ionization rates are more important, so that the
ionization state should depend mainly on
$C_i(T_e) \tau$.
For an under-ionized plasma where the electron temperature is greater than the ionization potential of
the relevant ions, $C_i(T_e)$ varies slowly,
$ C_i(T_e) \propto T_e^{1/2}$.
We have confirmed that this dependence fits the temperature dependence of the ionization rate of
\ion{O}{7} 
over  the interesting temperature range
($k T_e = $ 1--5~keV).
Below, we show that the ratio of \ion{O}{8} to \ion{O}{7} lines is the best diagnostic of departures from
ionization equilibrium
(Section~\ref{ion_sec:line}).

We assume that the electron temperature increases in proportion to the mean temperature $T_\Delta$
at the radius $R_\Delta$.
The mass within $R_\Delta$ is $M_\Delta = ( 4 \pi /3 ) R^3_\Delta \rho_\Delta$, so that the radius
$R_\Delta$ is given by
\begin{equation}
\label{ion_eq:RDelta}
R_\Delta
=
\left( \frac{2G}{H^2_0 \Delta} \right)^{1/3}
M_\Delta^{1/3}
[ E (z) ]^{-2/3}
\, .
\end{equation}
The condition of hydrostatic equilibrium or the shock jump condition at the accretion shock imply that
$ k T_\Delta \approx G M_\Delta / R_\Delta$.
Thus, the gas temperature varies as
\begin{equation}
\label{ion_eq:TDelta}
T_\Delta
\propto
M_\Delta^{2/3}
[ E (z) ]^{2/3}
\, .
\end{equation}
This implies that
\begin{equation}
\label{ion_eq:Cscale}
C_i (T_e) \propto
M_\Delta^{1/3}
[ E (z) ]^{1/3}
\, .
\end{equation}
Combining Equations~(\ref{ion_eq:tauscale}) and (\ref{ion_eq:Cscale}) gives
\begin{equation}
\label{ion_eq:Ctau}
C_i (T_e) \tau
\propto
M_\Delta^{1/3}
[ E (z) ]^{4/3}
\, .
\end{equation}
Equations~(\ref{ion_eq:tauscale}) and (\ref{ion_eq:Ctau}) suggest that the ionization state should depend only very weakly on the cluster mass, but should depend strongly on cluster redshift.
The increase of $\tau$ and $C_i (T_e) \tau$ with $z$ implies that 
collisional ionization will be faster at high redshifts, and hence
non-equilibrium ionization will be most important in low redshift clusters.

\begin{figure}
\includegraphics[angle=270,width=12cm]{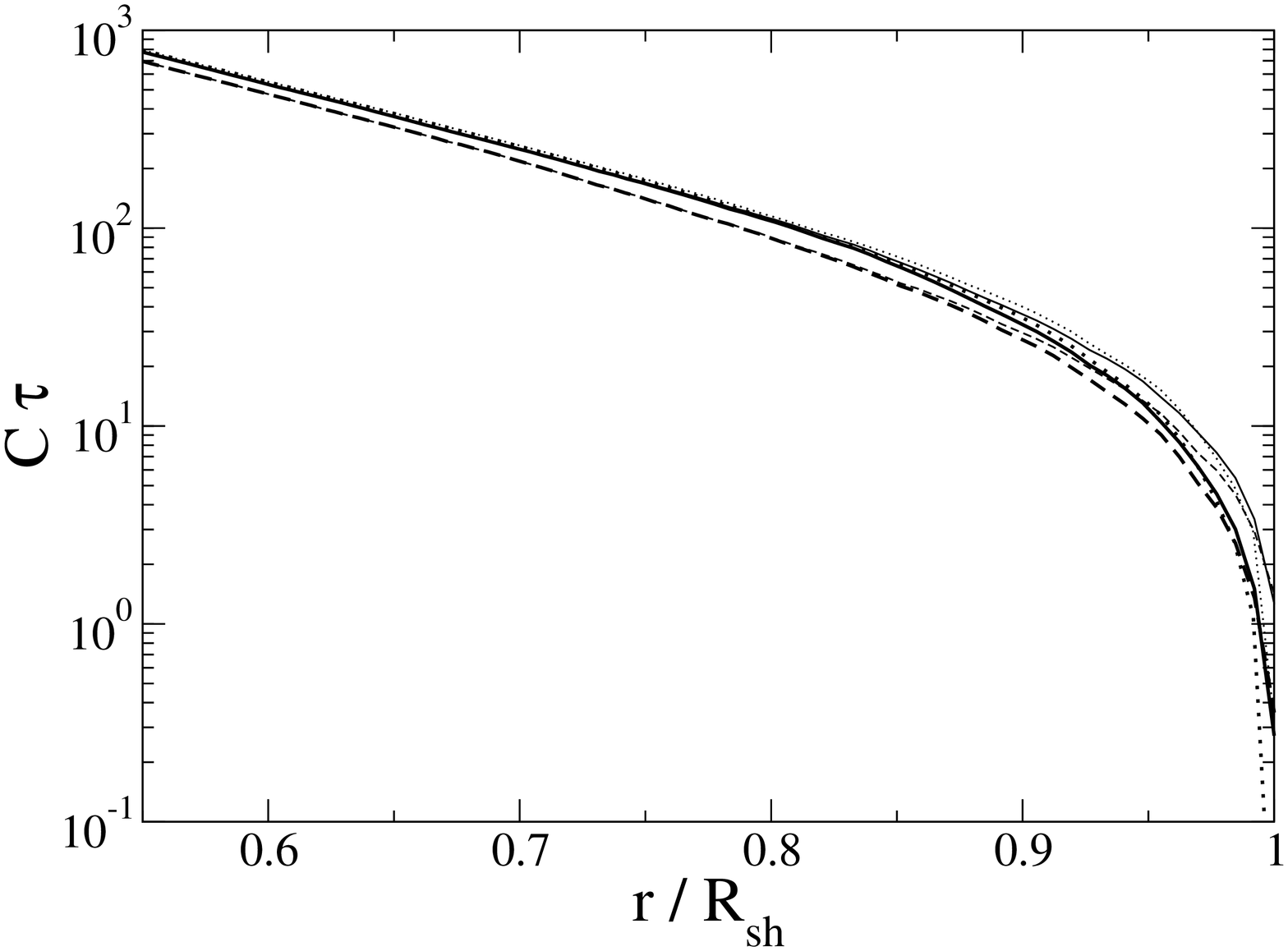}
\caption{
Dimensionless ionization parameter $C_i(T_e) \tau$ for \ion{O}{7}
versus the scaled radius ($r/R_{\rm sh}$) 
for clusters with total accreted
masses of $M_{\rm sh} = 0.77$ (dashed line), 1.53 (solid line), and 3.06
(dotted line) $\times$ $10^{15}~M_{\odot}$ at a redshift of $z=0$.
Models with $\beta = 1/1800$ (non-equipartition) and $\beta = 1$
(equipartition) are shown in thick and thin lines, respectively.
}
\label{ion_fig:Ctau_vs_m}
\end{figure}

Later, we will show that the ratio of \ion{O}{8} to \ion{O}{7} lines is the best diagnostic of departures from
ionization equilibrium
(Section~\ref{ion_sec:line}).
Figure~\ref{ion_fig:Ctau_vs_m} shows the dimensionless ionization   
parameter $C_i(T_e) \tau$ for \ion{O}{7} as a function of the scaled
radius ($r/R_{\rm sh}$) for clusters with different masses at $z=0$.
Here, $R_{\rm sh}$ is the radius of the cluster accretion shock.
Models with $\beta = 1/1800$ (non-equipartition) and $\beta = 1$
(equipartition) are shown in thick and thin lines, respectively.  All the
curves for the same $\beta$ but with different masses nearly overlap each
others.  This confirms that the non-equilibrium ionization effect at the
same characteristic radius is nearly independent of mass.

\begin{figure}
\includegraphics[angle=270,width=12cm]{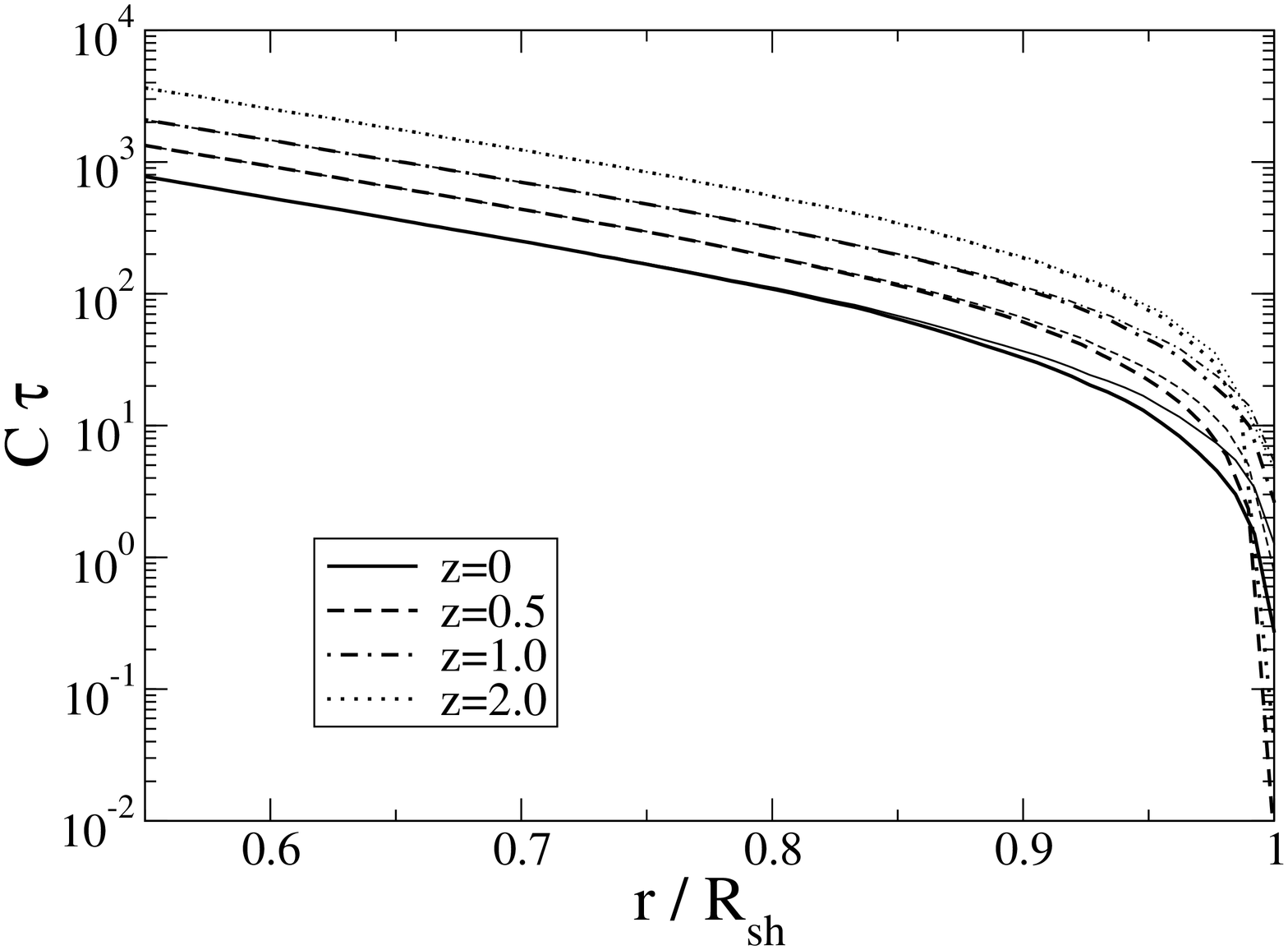}
\caption{
Dimensionless ionization parameter $C_i(T_e) \, \tau$ for \ion{O}{7} 
versus the scaled radius ($r/R_{\rm sh}$) for clusters at different redshifts.
The cluster model with a total accreted mass $M_{\rm sh} = 1.53 \times
10^{15}~M_{\odot}$ at $z=0$ is used.
Models with $\beta = 1/1800$ (non-equipartition) and $\beta = 1$
(equipartition) are shown in thick and thin lines, respectively.
}
\label{ion_fig:Ctau_vs_z}
\end{figure}

Figure~\ref{ion_fig:Ctau_vs_z} shows the dimensionless ionization
parameter as a function of the scaled radius for cluster models at different
redshifts.  The cluster model with a total accreted mass $M_{\rm sh} =   
1.53 \times 10^{15}~M_{\odot}$ at $z=0$ is used.  Models with $\beta =
1/1800$ (non-equipartition) and $\beta = 1$ (equipartition) are shown in
thick and thin lines, respectively.  In contrast to the dependence in
mass, we find that there is significant evolution with redshift.   The
non-equilibrium ionization effect should be strongest for low-redshift
clusters.

\begin{figure}
\includegraphics[angle=270,width=12cm]{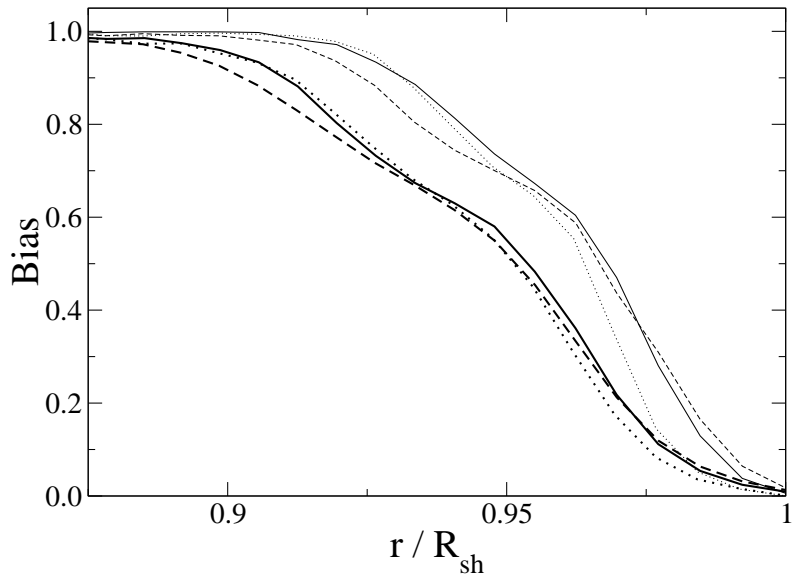}
\caption{
Non-equilibrium ionization bias versus the scaled radius ($r/R_{\rm sh}$) for 
clusters with total accreted masses of $M_{\rm sh} = 0.77$ (dashed line),
1.53 (solid line), and 3.06 (dotted line) $\times$ $10^{15}~M_{\odot}$ at
a redshift of $z=0$.
Models with $\beta = 1/1800$ (non-equipartition) and $\beta = 1$
(equipartition) are shown in thick and thin lines, respectively.
}
\label{ion_fig:OfracBias_vs_m}
\end{figure}

The preceding arguments assumed that departures from ionization equilibrium
could be assessed through the variation of $C_i(T_e) \tau$.
To test this explicitly and to determine if the preceding arguments apply to spectral diagnostics
for non-equilibrium ionization,
we study the effects of
non-equilibrium ionization on \ion{O}{7} and \ion{O}{8} ion fractions
using numerical simulations.  We calculate the non-equilibrium ionization
bias, $b$, of the ionization fraction ratios
$f($\ion{O}{8}$)/f($\ion{O}{7}$)$ for the 
non-equilibrium ionization (NEI) and collisional ionization 
equilibrium (CIE) 
models with the
same electron heating efficiency $\beta$.  The non-equilibrium ionization
bias is defined as
\begin{equation}
\label{ion_eq:bias}
b \equiv \frac{ \left[ f({\rm O~VIII})/f({\rm O~VII}) \right]_{\rm NEI} }
{ \left[ f({\rm O~VIII})/f({\rm O~VII}) \right]_{\rm CIE} }
\,.
\end{equation}
Figure~\ref{ion_fig:OfracBias_vs_m} shows the non-equilibrium ionization
bias as a function of the scaled radius for clusters with different masses.
Models with $\beta = 1/1800$ (non-equipartition) and $\beta = 1$
(equipartition) are shown in thick and thin lines, respectively.  The
nearly self-similar curves justify the semi-analytical argument that
non-equilibrium ionization effect at the same characteristic radius is
nearly independent of mass.  Note that the non-equilibrium ionization
effect on the $f($\ion{O}{8}$)/f($\ion{O}{7}$)$ ratio is only significant
for the outer 10\% of the shock radius.  However, the projected
\ion{O}{7} and \ion{O}{8} line emission will be significantly affected
even at a radius as small as one-fourth of the shock radius.   This is   
because the projected emission at the inner radius can be dominated by
the line emission from the under-ionized outer shell
(Section~\ref{ion_sec:line} below).

Figure~\ref{ion_fig:OfracBias_vs_z} shows the non-equilibrium ionization
bias as a function of the scaled radius for cluster models at different   
redshifts.  The cluster model with a total accreted mass $M_{\rm sh} =   
1.53 \times 10^{15}~M_{\odot}$ at $z=0$ is used.  Models with $\beta = 
1/1800$ (non-equipartition) and $\beta = 1$ (equipartition) are shown in  
thick and thin lines, respectively.  The non-equilibrium ionization effect is
larger for clusters at lower redshifts which agrees with the
semi-analytical argument given above.  Note that at a redshift higher than 1,
only a thin shell with a width of less than 5\% of the shock radius is in
non-equilibrium ionization compared to the wider shell with a width of
$\sim 10\%$ of the shock radius for zero redshift clusters.

\section{Non-Equilibrium Ionization Signatures}
\label{ion_sec:signatures}

\subsection{Models Used to Calculate Spectra}
\label{ion_sec:models}

Massive clusters at low redshifts are ideal candidates to study the 
non-equilibrium ionization effects in the outer regions since the departures from
ionization equilibrium are larger at low redshift and nearly independent of cluster mass
(Section~\ref{ion_sec:b_vs_mass_z}). 
Clusters with higher masses are 
more luminous in X-rays, and hence the spectral signatures are easier to detect. 
In the following, we present X-ray spectra for the 
hydrodynamic cluster model with an accreted mass of $M_{\rm sh} = 1.53 
\times 10^{15}~M_{\odot}$ at $z=0$ calculated in \citet{WS09}.  This 
model represents a typical massive cluster in the present-day  universe.
The shock radius is $R_{\rm sh} = 
4.22$~Mpc for this model.  The virial radius
is $R_{\rm vir} = R_{95}= 2.75$~Mpc, and the total mass 
within $R_{95}$ is $M_{95} = 1.19 \times 10^{15} M_{\odot}$.
Another commonly used radius and mass are $R_{200} = 1.99$~Mpc and $M_{200} = 
9.50 \times 10^{14} M_{\odot}$.

We calculate spectra for three different values of the shock electron heating
efficiency $\beta = 1/1800$, 0.5, and 1.
The last case corresponds to electron-ion equipartition.
We also calculate spectra both for 
non-equilibrium ionization (NEI) models and for collisional ionization equilibrium 
(CIE) models for comparison. 
For the NEI models, the results for the  model with the small shock electron heating efficiency
$\beta = 1/1800$ will be discussed extensively throughout the paper, as this model maximizes the
departures from equilibrium in the outer regions of clusters. 
For the CIE models, we present a 
non-equipartition model with $\beta = 1/1800$ (CIE--Non-Eq) and an 
equipartition model with $\beta = 1$ (CIE--Eq).

\subsection{X-ray Spectra}
\label{ion_sec:spectra}

The projected rest frame spectra for several different models at two projected radii are shown in 
Figure~\ref{ion_fig:spec}.  In the upper panels, we show the NEI model 
with a very small shock heating efficiency ($\beta = 1/1800$), and 
electrons and ions are in non-equipartition.  The CIE--Non-Eq model is 
shown in the lower panels.  The left panels show spectra at a radius of 
$r=2$~Mpc, while the right panels show spectra at a radius of 
$r=3.5$~Mpc.  Each spectrum is binned with a bin size of $\Delta \log 
(E)=0.005$.  

At $r=2$~Mpc, the overall spectra are dominated by the free-free 
continuum emission over a wide range of energy for both the NEI and the 
CIE--Non-Eq models.  The continuum spectra for both models are nearly 
identical because of the dominant free-free emission with the same 
electron temperature.
The line emission is also very similar for both models because at this 
radius, the ionization timescale parameter is rather large
($\tau \sim 8 \times 10^{12}$~cm$^{-3}$~s).  The most notable difference 
is the line intensity of the \ion{O}{7} triplet lines near $\sim 
0.57$~eV.  This line intensity for the NEI model is much higher than that 
of the CIE--Non-Eq model.  There is almost no \ion{O}{7} line emission 
for the CIE--Non-Eq model.  The strong \ion{O}{7} at $r=2$~Mpc for the 
NEI model is mainly due to the projection of emission from the 
under-ionized \ion{O}{7} ions in the outer regions with much shorter 
ionization timescale parameters.  The differences for other strong emission lines
are much smaller between the two models.  There is slightly more soft 
line emission below about 1~keV for the NEI model.

At $r=3.5$~Mpc, there are significant differences in the spectra between 
the NEI and the CIE--Non-Eq models.  For the CIE--Non-Eq model, the 
continuum emission still dominates the overall spectrum; for the 
NEI model, the soft emission is dominated by lines.  One of 
the most obvious differences in the line emission between the two models 
is again for the \ion{O}{7} triplet.  The line intensity for the NEI 
model 
is much higher than that of the CIE--Non-Eq model.  The \ion{O}{7} 
triplet is weak for the CIE--Non-Eq model.  By inspecting a 
number of line ratios at different radii, we found that the ratio of 
the \ion{O}{7} and \ion{O}{8} line intensities can be used as a 
diagnostic for the degree of ionization equilibrium, and this will be 
discussed in Section~\ref{ion_sec:line} below.

\subsection{Surface Brightness Profiles}
\label{ion_sec:sb}

The rest frame radial surface brightness profiles integrated over 
various energy 
bands for the outer regions of clusters are shown in 
Figure~\ref{ion_fig:sb}.  The NEI models with $\beta = 1/1800$ are shown 
in thick lines in both the upper left and upper right panels.  The 
CIE--Non-Eq and the CIE--Eq models are shown as thin lines on the upper 
left and upper right panels, respectively.  The ratios of $S_{\rm NEI}/ 
S_{\rm CIE-Non{\text -}Eq}$ and $S_{\rm NEI}/ S_{\rm CIE-Eq}$ are shown 
below the corresponding panels.  Comparing the NEI and the CIE--Non-Eq 
models tells us the effects of non-equilibrium ionization alone, while 
comparing the NEI and the CIE--Eq model tells us the total effects of 
both non-equilibrium ionization and non-equipartition.

From the left panels of Figure~\ref{ion_fig:sb}, we can see that 
non-equilibrium ionization significantly enhances the soft (0.3--1.0~keV) 
emission in the outer regions.  For the NEI model, the soft emission 
has been increased by more than 20\% at around 3~Mpc, and up to nearly an 
order of magnitude around the shock radius compared to the CIE--Non-Eq 
model.  The increase of the soft emission is due to the line emission 
by the under-ionized ions.  For the CIE--Non-Eq model, the surface 
brightness profiles in all energy bands shown in Figure~\ref{ion_fig:sb} 
decrease rapidly out to the shock radius.  For the NEI model, the 
decrease in surface brightness in the soft band as a function of radius 
slows down near $\sim 3$~Mpc, and the soft band surface brightness actually increases with radius
from $\sim$3.7~Mpc out to nearly the shock radius, where the surface brightness drops rapidly.
Within about 2.3~Mpc, non-equilibrium ionization 
effect is less than 5\% in the soft band.  The non-equilibrium ionization 
effect on the medium (1.0--2.0~keV) band is not as dramatic as the soft 
band, and the maximum increase is about 70\% near 3.8~Mpc.  The 
non-equilibrium ionization effect on the hard (2.0--10.0~keV) band is 
less than 5\% in most regions, and the effect is to lower the hard 
emission near the shock radius.   The soft emission dominates the 
overall X-ray band (0.3--10.0~keV) outside of $\sim$3~Mpc, and the overall 
X-ray emission decreases more slowly than for the CIE--Non-Eq model 
beyond that radius.  Near the shock radius, the surface brightness in the 
overall X-ray band for the NEI model is about a fact of 6 higher than 
that for the CIE--Non-Eq model.

The right panels of Figure~\ref{ion_fig:sb} show that in addition to the 
non-equilibrium ionization effect, non-equipartition will increase the soft
emission by a significant factor  near the shock radius.  This occurs
because in the CIE--Non-Eq and NEI models, the 
electron temperatures in the outer regions are lower than for the 
CIE--Eq model, and this also leads to more soft X-ray line emission.
The surface brightness profile in the overall 
X-ray band for the NEI model is a factor of 5 higher than that for the 
CIE--Eq model near the shock radius.  A detailed discussion on the 
difference between the CIE--Non-Eq and CIE--Eq models can be found in 
\citet{WS09}.

\subsection{\ion{O}{7} and \ion{O}{8} Line Ratio}
\label{ion_sec:line}

The most prominent non-equilibrium ionization signature in the X-ray lines 
are for the line ratio of \ion{O}{7} and \ion{O}{8}.  
Figure~\ref{ion_fig:spec_line} shows the spectra for models at $r=2$~Mpc.  
The spectra are shown in the 0.5--0.7~keV range which covers the rest 
frame energies of the \ion{O}{7} and \ion{O}{8} lines.  The upper 
panel shows the NEI model with a very low electron heating efficiency 
$\beta=1/1800$.  The middle panel shows the NEI model with an 
intermediate $\beta=0.5$.  The lower panel shows the CIE--Non-Eq model 
with $\beta=1/1800$.  Each spectrum is binned with a bin size of $\Delta 
E=0.0001$~keV.

In the upper panel of Figure~\ref{ion_fig:spec_line}, the most prominent 
lines are the He-like \ion{O}{7} triplets at 561.0, 569.6, and 
574.0~eV, the H-like \ion{O}{8} doublets at 653.5 and 653.7~keV, and 
the He-like \ion{O}{7} line at 665.6~eV.  Here, we focus on the line 
ratio between the He-like \ion{O}{7} triplets and the H-like \ion{O}{8} 
doublets as a diagnostic for non-equilibrium ionization. 
These two lines have been used to search for the WHIM as well as to study 
the non-equilibrium ionization of the WHIM \citep{CF06,YS06}.  We 
find that the He-like \ion{O}{7} and the H-like \ion{O}{8} lines 
also show strong signatures of non-equilibrium ionization in the outer
regions of clusters.

The upper and lower panels of Figure~\ref{ion_fig:spec_line} show that 
non-equilibrium ionization strongly enhances the He-like \ion{O}{7} 
triplets compared to the CIE--Non-Eq model.  The H-like \ion{O}{8} 
doublets for the NEI model are only slightly stronger than for the 
CIE--Non-Eq model.  This suggests that the ratio of the \ion{O}{7} and 
\ion{O}{8} lines can be used as a diagnostic for
non-equilibrium ionization.  The \ion{O}{7} and \ion{O}{8} lines are 
similarly strong for the NEI models with electron 
heating efficiencies $\beta=0.5$ and $1/1800$, and we suggest that the 
\ion{O}{7} and \ion{O}{8} line ratio is a good diagnostic for the 
non-equilibrium 
ionization for a wide range of electron 
heating efficiencies.

Figure~\ref{ion_fig:line_flux} shows the surface brightness for the 
\ion{O}{7} and 
\ion{O}{8} lines for the NEI model with $\beta = 1/1800$.  The surface 
brightness of the lines are calculated by subtracting the continuum 
surface brightness from the 
total surface brightness within narrow energy ranges of 556.0--579.0~eV 
and 
648.5--658.7~eV for the \ion{O}{7} triplets and the \ion{O}{8} 
doublets, respectively.  The energy bands were chosen to cover the 
\ion{O}{7} and the \ion{O}{8} lines with a spectral resolution of
10~eV which is the expected value for the outer arrays of the 
X-ray Microcalorimeter Spectrometer (XMS) 
on the {\it IXO}. 
The continuum surface brightness were calculated by 
fitting the spectrum to a power law model in the energy range of 
0.5--0.7~keV, excluding the lines.
This simulates the techniques likely to be used to analyze real observations. 
We also show the continuum surface brightness within the narrow 
energy bands used to extract the line surface brightness in 
Figure~\ref{ion_fig:line_flux}.

The surface brightness of the \ion{O}{7} triplets is nearly constant 
from $\sim 1$~Mpc to $\sim 3$~Mpc, and then rises gradually.
Note that a flat surface brightness at inner radii and rising surface 
brightness at larger radii is the signature of a shell of emission at 
large radii seen in projection.
That is, most of the \ion{O}{7} emission is actually at large radii 
where the ionization time scale is short.
Beyond about 3.9~Mpc, 
the \ion{O}{7} surface brightness rises rapidly to a peak value, and 
then drops.  
The continuum in the 556.0--579.0~eV energy band drops from a radius of 
1~Mpc out to the shock radius.  Beyond $\sim 2.8$~Mpc, the \ion{O}{7} 
line emission dominates over the continuum in the 556.0--579.0~eV energy 
band.  For the \ion{O}{8} doublets, the surface brightness drops from a 
radius 
of 1~Mpc out to $\sim 3$~Mpc, and then rises to a peak value at $\sim 
4$~Mpc.  The \ion{O}{8} surface brightness then drops beyond $\sim 
4$~Mpc.  
The \ion{O}{8} line emission dominates over the continuum in the 
648.5--658.7~eV energy band beyond $\sim 2$~Mpc.

Figure~\ref{ion_fig:lineratio} shows the line ratios of \ion{O}{7} and 
\ion{O}{8}, $S($\ion{O}{8}$)/S($\ion{O}{7}$)$, for different 
models.  To compare the effect of non-equilibrium ionization alone, we 
can compare the line ratios between the NEI model (solid line) and the 
CIE--Non-Eq model (dashed line), while both models assume an electron 
heating efficiency $\beta = 1/1800$.  Both models use the same 
non-equipartition electron temperature to calculate the spectra, but one 
of them assumes equilibrium ionization while the other one does not.  At 
$\sim 1$~Mpc, the line ratio for the NEI model is more than a factor of 2 
lower than that of the CIE--Non-Eq model.  The difference increases as 
the radius increases, and the differences are over an order of magnitude 
for radii beyond $\sim 2$~Mpc.  For the CIE--Eq model (dotted line), the 
line ratio is very similar to that of the CIE--Non-Eq model, except near 
the shock regions where the line ratio for the the CIE--Eq model is a 
factor of a few higher than the CIE--Non-Eq model.  Both the line ratios 
of the CIE--Eq and the CIE--Non-Eq models are above 10 in most regions 
between 1--4~Mpc.  NEI models with electron heating efficiencies $\beta = 
0.5$ and $1.0$ are also shown in Figure~\ref{ion_fig:lineratio}.  The
effect of increasing the electron heating efficiency is to raise the 
electron temperature
at the shock, and hence increase the ionization rates. 
From Figure~\ref{ion_fig:lineratio}, we can see that increasing the 
electron heating efficiency only affects the line ratio by less than a 
factor of two in most regions shown compared to the NEI model with $\beta 
= 1/1800$.  The line ratios for all the NEI models with different 
electron heating efficiencies are less than 10 for radii beyond
about 1.3 Mpc.  In summary, the line ratios in the outer regions for all the 
NEI models we calculated are significantly smaller than those for the CIE 
models, and the differences are larger than an order of magnitude for 
most regions beyond $\sim 2$~Mpc.  Such large differences can be used to 
distinguish between NEI and CEI models in real observations.

\section{Detectability of \ion{O}{7} and \ion{O}{8} Lines with {\it IXO} 
and Testing the Non-Equilibrium Ionization Effect}
\label{ion_sec:detect}

In this section, we estimate whether the \ion{O}{7} and \ion{O}{8} 
lines in cluster outer regions
and the non-equilibrium ionization signatures 
can be detected with {\it IXO}.
We consider the NEI model with 
$\beta = 1/1800$  for a cluster with
an accreted mass of $M_{\rm sh} = 1.53 
\times 10^{15}~M_{\odot}$ at low redshift.
The choice is justified by the fact that the non-equilibrium ionization 
effect does not depend strongly on mass and the effect is larger at lower 
redshift (Section~\ref{ion_sec:b_vs_mass_z}), and that the surface 
brightness for massive and low redshift clusters is higher.

The X-ray Microcalorimeter Spectrometer (XMS) planned for the {\it IXO} 
can potentially detect the \ion{O}{7} and \ion{O}{8} lines in 
cluster outer regions.  
The Wide Field Imager (WFI) does not have enough spectral resolution 
($>50$~eV) 
at around 0.6~keV, and the \ion{O}{7} triplets cannot be resolved from 
the 0.5~keV nitrogen line.  The X-ray Grating Spectrometer covers the 
interesting energy range, but the collecting area is too small to detect 
the weak \ion{O}{7} and \ion{O}{8} lines.  Therefore, we only 
consider the XMS in our estimations.  The XMS has inner (core) and outer
microcalorimeter arrays with expected spectral resolutions of 2.5 and 
10~eV, respectively.  We consider two cases when observing with the XMS.  
The first case (XMSC) is that only the inner core array is used, and 
the second case (XMSF) is that the full array (both inner and outer 
arrays) is used.
For simplicity, when considering 
the XMSF, we assume the spectral resolution to be the same as the outer 
array throughout the full array.
This doesn't strongly affect the detectability of the lines, since the 
bands used to determine the fluxes in the lines are set by the line width 
in the outer array, and the line width in the inner core doesn't affect 
the line flux.
The expected effective areas ($A_{\rm eff}$) 
for the XMS is about 10000~cm$^2$ at around 0.6~keV.  The relevant 
instrument parameters for the two cases we considered are listed in 
Table~\ref{ion_tab:IXO}.

\subsection{Backgrounds}
\label{ion_sec:bg}

The major background for the {\it IXO} observations are Non-X-ray 
Background 
(NXB), the soft emission from the local Galactic background (GXB), and 
these are included in our signal-to-noise ratio calculations.  The 
cluster continuum emission will be much weaker than the line emission 
in the band widths we are interested in, but we also include the cluster 
continuum emission in our calculations.  With the spatial resolution of 
5\arcsec\ and the very long exposure time needed for the line detection, 
point source contaminations will be negligible, and hence it is not 
included in our calculations.  In fact, the GXB we used has included a 
component from unresolved AGNs, and this may overestimate the total 
background.  
The total count rates of the backgrounds we used for the two instrument 
setups are listed in Table~\ref{ion_tab:bg}, and are discussed below.

We use the GXB simulated by \citet{FCS+05} which included two thermal 
components to represent the Local Hot Bubble emission and the 
transabsorption emission \citep{Sno98, KS00}, and one continuum emission 
component to represent unresolved AGN background.  The parameters used 
for their background model are based on \citet{McC+02}.  The most 
prominent emission around 0.6~keV is the line emission from nitrogen 
and oxygen ions \citep[Figure~6 in][]{FCS+05}.  With the very high XMS 
spectral resolutions, the \ion{O}{7} and \ion{O}{8} lines from 
clusters beyond $z\approx 0.028$ should be separated from the strong line 
emission in the GXB (see Table~\ref{ion_tab:min} below).  We adopt the 
continuum intensity value of $I_{\rm GXB} = 24$ 
photons~cm$^{-2}$~s$^{-1}$~sr$^{-1}$~keV$^{-1}$ at around 0.6~keV, which 
is used in \citet{FCS+05}.  The total count rate for the whole field of 
view ($\Omega_{\rm FOV}$) which covers the energy range of the lines is 
given by $R_{\rm GXB} = I_{\rm GXB} \times A_{\rm eff} \times \Omega_{\rm 
FOV} \times \Delta E_{\rm band}$, where $\Delta E_{\rm band}$ is the 
bandwidth covered the lines.  For the \ion{O}{8} doublets, since the 
line separation is much smaller than the spectral resolutions of all the
instruments we considered, $\Delta E_{\rm band}$ is simply the spectral 
resolution of the corresponding instrument.  For the \ion{O}{7} 
triplets, the three lines should be well separated by the XMS core array.  
Hence, $\Delta E_{\rm band}$ of \ion{O}{7} is then three times the 
spectral resolution of the XMS core array.  However, for the XMS full 
array, the \ion{O}{7} triplets cannot be resolved.  Therefore, $\Delta 
E_{\rm band}$ is then given by the maximum separation of the triplets 
(13~eV) plus the corresponding spectral resolutions.

For a future mission like the {\it IXO}, the NXB is rather uncertain.  
We use the count rate of $F_{\rm NXB} = 8.1 \times 10^{-3}$ 
photons~s$^{-1}$~arcmin$^{-2}$~keV$^{-1}$ at 0.6~keV for the XMS 
estimated by the {\it IXO} team \citep{Smi08}.
The total count rate for the 
whole field of view which covers the energy range of the lines is given 
by $R_{\rm NXB} = F_{\rm NXB} \times \Omega_{\rm FOV} \times \Delta 
E_{\rm band}$.  
To address the effects of the uncertainties, we also multiply the NXB by 
factors of 0.5 and 2 in our calculations 
(Section~\ref{ion_sec:SN} below).

The cluster continuum emission within the very narrow energy bands we 
are interested in are much weaker than the line emission in cluster 
outer regions (Figure~\ref{ion_fig:line_flux}) and the GXB.  To detect 
the \ion{O}{7} and \ion{O}{8} lines, it is also important to 
observe regions where the cluster continuum emission is weak compared 
to the line emission.  Therefore, the cluster continuum emission should 
not be important when estimating the signal-to-noise ratio.  
Nevertheless, we have included the cluster continuum emission in our 
calculations.  To be conservative, we simply take the cluster continuum 
emission at $r=2.8$~Mpc where the cluster continuum emission across 
648.5--658.7~eV (energy range where the \ion{O}{7} triplets are 
covered by the XMS full array) equal to the \ion{O}{7} emission.  The 
continuum count rates $R_{\rm cont}$ are listed in 
Table~\ref{ion_tab:bg}.

\subsection{Signal-to-Noise Ratio}
\label{ion_sec:SN}

In order for the lines to be separated from the local GXB, clusters to be 
observed should be at high enough redshifts.  Since the \ion{O}{7} 
triplets spread across 13~eV, the best targets should have redshifts such 
that the lines can be shifted by at least 13~eV plus the spectral 
resolution.  The targets should also not to have too high redshift such 
that the \ion{O}{7} triplets will not overlap with the 0.5~keV nitrogen 
line.  Table~\ref{ion_tab:min} lists the minimum and maximum redshifts 
for clusters to be observed by the XMS core and the XMS full arrays which 
meet these requirements.

To calculate the signal of the \ion{O}{7} and \ion{O}{8} lines for 
clusters for both the XMSC and XMCF cases, we assume the nominal massive 
cluster model to be at a redshift of 0.05.  The redshift is chosen such 
that it is slightly higher than $z_{\rm min}$ for the XMS full array in 
Table~\ref{ion_tab:min}.
It is also important that there actually be clusters which are at or within the selected redshift  which are 
fairly regular in shape in their outer regions and with  masses which are comparable to the cluster model we calculated (e.g., Abell 85, Abell 1795).

To calculate the total count rates of the lines from a cluster at the
assumed redshift as observed by {\it IXO} ($R_{\rm line}$), we first 
convolve the 
rest frame surface brightness of the lines 
(Figure~\ref{ion_fig:line_flux}) with a 
top hat function with a width of 0.114 (0.286)~Mpc which corresponds to 
the physical distance at $z=0.05$ covered by the XMSC (XMSF) field of 
view, and then multiply the
convolved rest frame surface brightness by a 
factor of 
$A_{\rm eff} \Omega_{\rm FOV} (4\pi)^{-1} (1+z)^{-3}$ to give the count 
rate $R_{\rm line}$.  
In calculating the signal-to-noise ratios, count rates at a radius where 
the surface brightness of the weaker line is maximum in the outer regions 
are used as the optimum model count rates.  The optimum count rates are 
also listed in Table~\ref{ion_tab:bg}.

We calculate the signal-to-noise ratios for the optimum models (1 NXB, 1 
GXB) which use the count rates listed in Table~\ref{ion_tab:bg}.
The signal for each line is given by
$ S_{\rm ctn}  = R_{\rm line} \, t $, 
where $t$ is the exposure time.
The noise for each line is given by
\begin{equation}
\label{ion_eq:noise1}
N = \sqrt{ ( R_{\rm line} + R_{\rm cont} + R_{\rm GXB} + 
R_{\rm NXB} ) \, t } \,.
\end{equation}
The signal-to-noise ratio of the line ratio is given by
\begin{eqnarray}
\label{ion_eq:lineratio}
SN({\rm O~VIII}/{\rm O~VII}) = &
\nonumber \\ 
& \hskip -28mm
\left[ (SN({\rm O~VIII}))^{-2} + (SN({\rm 
O~VII}))^{-2} \right]^{-1/2} \,,
\end{eqnarray}
where $SN({\rm O~VII})$ and $SN({\rm O~VIII})$ are the signal-to-noise 
ratios of the \ion{O}{7} and \ion{O}{8} lines, respectively.

To address the effects of the NXB uncertainties, we also calculate the 
signal-to-noise ratios for models with the NXB multiplied by a factor of 
2 (2 NXB, 1 GXB) and a factor of 0.5 (0.5 NXB, 1 GXB).  The GXB varies 
with sky position.  To address the uncertainties in the GXB, we vary the 
GXB by a factor of 0.8 (1 NXB, 0.8 GXB), 1.2 (1 NXB, 1.2 GXB), and 1.5 (1 
NXB, 1.5 GXB).  In real observations, the line signals may not be 
optimum.  We consider models with the count rates of the lines to be a 
factor of 0.75 (0.75 $R_{\rm line}$), 0.50 (0.50 $R_{\rm line}$), and 
0.25 (0.25 $R_{\rm line}$) of the optimum models.

The signal-to-noise ratios as a function of exposure time for the 
different models are shown in Figure~\ref{ion_fig:SN1}.
To ensure that there 
are enough counts for the lines, we only plot the signal-to-noise ratios 
if the total count for each line $S_{\rm ctn}$ is larger than 30.

For the XMSC instrument setup with the optimum model 
(1 NXB, 1 GXB), in order to get a signal-to-noise ratio of 3 for the 
\ion{O}{7} triplets, about 220~ksec is needed.  With a deeper 
observation of 600~ksec, the signal-to-noise ratio can exceed 5.  
For the \ion{O}{8} 
doublets, a shorter exposure time of $\sim 180$~ksec is need to 
have $ S_{\rm ctn}  >30$ with a signal-to-noise ratio of 3.2.  About 
450~ksec is needed to get a signal-to-noise ratio of 5, and it is 
possible to get a signal-to-noise ratio up to 7 with a very deep 
observation of 900~ksec.  
If no single cluster is observed for such a 
long exposure, the very deep exposure can be achieved by stacking many 
observations of outer regions of many clusters.  In this case, the line 
emission measured is average over many clusters.  
For a given exposure 
time, increasing the NXB by a factor of 2 or increasing the GXB by a 
factor of 1.5 only decrease the signal-to-noise ratios by about 10\% for 
both the \ion{O}{7} or \ion{O}{8} lines.  As expected, 
decreasing 
the line signals significantly decreases the signal-to-noise ratios.  If 
the line signals are smaller than half of the optimum values, it 
will be impossible to detect the \ion{O}{7} (\ion{O}{8}) line with an 
exposure time less than 
$\sim 0.8$ (0.5)~Msec.
Limited by the \ion{O}{7} detection, an 
exposure time of about 220~ksec is sufficient to have a 2.3-$\sigma$ 
determination for the line ratio for our optimum model.
A 3-$\sigma$ measurement of the line ratio will require an exposure time 
of about 380~ksec.

If the electron temperature can be measured from observations such as 
X-ray spectroscopy or hardness ratios, the measured line ratio can be 
compared to the CIE ratio inferred by the electron temperature, 
and this will provide a direct test to the ionization states of the plasma 
without ambiguity.  In the outermost regions of a cluster, the surface 
brightest may be too low and hence reliable temperature measurement may 
not be feasible.   However, as shown in Figure~\ref{ion_fig:lineratio}, 
beyond a radius of $\sim 2.5$~Mpc where the electron temperatures are 
always higher than about 0.5~keV for either the non-equipartition model 
with $\beta = 1/1800$ or the equipartition model, the CIE line ratios are 
always higher than $\sim 4$.  In contrast, the NEI line ratios are always 
lower than $\sim 2$ for the electron temperature range of interest in 
realistic 
clusters for all the models we considered.  Hence, a 3-$\sigma$ 
measurement of the line ratio $\lesssim 2$ ($\Delta 
[S($\ion{O}{8}$)/S($\ion{O}{7}$)] \lesssim$ 2/3 at 1-$\sigma$) is a 
strong evidence that 
the ions are in non-equilibrium ionization due to the huge difference 
between the NEI and CIE line ratios for the temperature range of realistic 
clusters.  
If the line ratio is measured to be even lower (e.g., $\lesssim 1$), only 
a 2 or 3-$\sigma$ will be sufficient to rule out the CIE model.  
On the other hand, if the line ratio is measured to be as high as 4 in the 
outermost regions, a 6-$\sigma$ ($\Delta 
[S($\ion{O}{8}$)/S($\ion{O}{7}$)$] = 2/3 at 1-$\sigma$) will be necessary 
to rule out the NEI low line ratio of $\sim 2$ at a 3-$\sigma$ level.
Even if the line ratio is measured within about $\lesssim 
2$~Mpc where the line ratio for the NEI model can be as high as $\sim 4$, 
the surface brightness at this relative small radius is high enough so 
that electron temperature can be measured by the {\it IXO} or even 
{\it Suzaku} 
\citep[e.g.,][]{GFS+09,Rei+09, Bas+10,Hos+10} with confidence.  To study 
whether the {\it IXO} XMS can measure the electron temperature at this 
radius, 
we have carried out simulations using XSPEC with our model spectrum taken 
at radius $\sim 2$~Mpc.  The detector response and background files were 
taken from the latest {\it IXO} simulator 
SIMX\footnote{http://ixo.gsfc.nasa.gov/science/simulator.html}.  Both the 
NXB and the GXB have been included in the background file.  We assumed an 
absorption model {\it wabs} with $n_H = 5 \times 10^{20}$~cm$^{-2}$.  We 
found that 
the {\it IXO} XMS core detector can collect about 2400 photon counts 
(0.2--3.0~keV) for a 150~ksec observation.  When fitting the 
continuum of the spectrum, we assumed an absorbed thermal model ({\it 
wabs*apec}) and ignored the strong lines.  Using the {\it wabs*nei} model, 
the temperature determined is within the uncertainty of the {\it 
wabs*apec} model and does not affect the results significantly.
We found that the electron 
temperature can be constrained to $T_e = 4.0^{+1.6}_{-1.5}$~keV.  The CIE 
line ratios within the determined temperature range are always higher than 
13, and hence a 2 to 3-$\sigma$ measurement for the low line ratio will 
also be enough to test the non-equilibrium ionization model.  In fact, the 
{\it IXO} WFI should constrain the electron temperature better because of 
its lower particle background \citep{Smi08}.

For the XMSF instrument setup, with the optimum model (1 NXB,
1 GXB), about $\sim 130~(100)$~ksec is needed to get a signal-to-noise 
ratio of 3 for the \ion{O}{7} triplets (\ion{O}{8} doublets).  A 
longer exposure of $\sim 350~(270)$~ksec is needed  to get a 
signal-to-noise ratio of 5 for the \ion{O}{7} (\ion{O}{8}) lines, 
and a very deep exposure of $\sim 1~(1)$~Msec is needed for a 
8~(10)-$\sigma$ detection.  
Increasing the NXB by a factor of 2 or increasing the GXB by a factor of 
1.5 decreases 
the signal-to-noise ratios by less than $\sim 13\%$ for both the 
\ion{O}{7} and \ion{O}{8} lines.  
Similarly to the XMSC instrument setup, 
the signal-to-noise ratios decrease as the line signals decrease.  
For the optimal model, a shorter exposure time of about 130~ksec (also 
limited by the \ion{O}{7} detection) compared to the XMSC setup will give 
the same 2.3-$\sigma$ detection for the line ratio.  
An exposure time of about 230~ksec will be needed for a 3-$\sigma$ 
measurement 
of the line ratio.

Overall, the XMSF instrument setup can achieve higher signal-to-noise 
ratios for a given exposure time compared to the XMSC setup for the 
optimum models.  This is mainly due to the larger field of view of the 
detector which can collect more photons with the same exposure time.  
The XMSF is slightly more subject to the continuum background 
uncertainties (cluster continuum emission, NXB, and GXB) due to the 
poorer spectral resolution.  For both the XMSC and XMSF setups, detecting 
the \ion{O}{7} and \ion{O}{8} lines
and testing non-equilibrium ionization
in cluster outer regions are promising.

\section{Discussions and Conclusions}
\label{ion_sec:conclusion}

Studying the physics in the outer regions of clusters is very important to 
understand how clusters are formed, how the the intracluster gas is
heated, as well as to constrain the formation of large-scale structure.
Because of the very low density in cluster outer 
regions, the collisional timescales are very long and comparable to 
the cluster age.  Electrons and ions passed through the accretion shocks 
may not have enough time to reach equipartition and the ions may be 
under-ionized.  In a previous paper \citep{WS09}, we have studied the 
non-equipartition effects on clusters using one-dimensional hydrodynamic 
simulations.  In this paper, we systematically studied non-equilibrium 
ionization effects on clusters and the X-rays signatures using the same 
set of simulations we have developed \citep{WS09}.

By using semi-analytic arguments together with numerical simulations, we 
have shown that the non-equilibrium ionization effect is nearly 
independent of cluster mass but depends strongly on redshift.  In particular, 
non-equilibrium ionization effects are stronger for low-redshift 
clusters.  Therefore, the brighter 
massive clusters at low-redshifts are good candidates for studying the 
non-equilibrium ionization effects.

We systematically studied non-equilibrium ionization signatures in X-rays 
for a 
massive cluster with $M_{\rm sh} = 1.53 \times 10^{15}~M_{\odot}$.  We 
first calculated the ionization fractions for 11 elements heavier than He 
following the electron temperature and density evolutions of each fluid 
element.  We then calculated the X-ray emissivity of each fluid element 
and the resulting projected spectra for the cluster.  Since the electron 
temperature profiles depend on electron heating efficiency $\beta$,
we have considered three 
different possibilities which represent a very low heating efficiency 
($\beta = 1/1800$), an intermediate heating efficiency ($\beta = 0.5$), 
and equipartition, $\beta = 1$.
We also considered models which assume equilibrium ionization for 
comparison.

At a radius (e.g., 2~Mpc) where the ionization timescale is long, the 
overall spectra for the NEI and CIE--Non-Eq models are very similar.  
This is because of the dominant free-free emission, and both models 
assume the same electron temperature.  However, in the outer regions, 
e.g, at $r \sim 3.5$~Mpc which is between $R_{\rm vir}$ and $R_{\rm 
sh}$, the soft emission in the NEI model is dominated by line 
emission, where the CIE--Non-Eq spectrum is still dominated by the 
continuum free-free emission.  

By analyzing the surface brightness profiles, we found that soft 
emission (0.3--1.0~keV) for the NEI model can be enhanced by more than 
20\% at around 3~Mpc, and up to nearly an order of magnitude near the 
shock radius compared to the CIE--Non-Eq model. The soft emission 
enhancement is mainly due to the line emission from under-ionized ions.  
The non-equilibrium ionization effects on the medium (1.0--2.0~keV) and 
hard (2.0--10.0~keV) band emissions are smaller.  The overall X-ray band 
(0.3--10.0~keV) emission is dominated by the soft emission, and the 
total X-ray emission for the NEI model decrease much slower than that of 
the CIE--Non-Eq model.  Thus, if cluster outer regions are in 
non-equilibrium ionization, the shock region will be much more luminous 
compared to the CIE--Non-Eq model.  

By inspecting a number of spectra, we found that the most prominent 
non-equilibrium ionization signature in line emission is the line ratio 
of the 
He-like \ion{O}{7} triplets and the H-like \ion{O}{8} doublets, 
$S($\ion{O}{8}$)/S($\ion{O}{7}$)$.  The line ratios for the CIE models 
are higher than 10 for most regions between $r=$~1--4~Mpc, while the line 
ratios are smaller than 10 for the NEI models.  The differences in the 
line ratios between the NEI and CIE models increase with radius, and the 
differences are more than an order of magnitude for radii beyond $\sim 
2$~Mpc.  These results are insensitive to the degree of non-equipartition 
or electron heating efficiency $\beta$.  We suggest that the line ratios 
can be used to distinguish between the NEI and CIE models.
The electron temperature profile can be determined from fits to the 
continuum spectra of the outer regions of clusters, allowing the CIE line 
ratios to be determined.  Comparison to the observed ratios should show 
the effects of non-equilibrium ionization.
Note that a line ratio of $S($\ion{O}{8}$) / S($\ion{O}{7}$) < 3$ in the 
outer region of a massive clusters is a clear signal of NEI.

We have also studied the detectability of the \ion{O}{7} and \ion{O}{8} 
lines around cluster accretion shock regions with {\it IXO}, as well as 
the test for non-equilibrium ionization using the line ratio.  For our 
optimum model, we found that with the XMS core array, an exposure time of 
220~ksec is need to have a 3.0-$\sigma$ detection of the \ion{O}{7} lines 
and about 180~ksec is need to have $>30$ counts for a 3.2-$\sigma$ 
detection of the \ion{O}{8} lines.  The uncertainties in NXB and GXB will 
not affect the results significantly.  For the XMS full array while we 
assume the spectral resolution to be the same as the outer array 
throughout the detector, we found that the signal-to-noise ratios for our 
optimum model are higher for the same exposure time as the XMS core 
array.  In particular, only about 130 (100)~ksec is needed to detect the 
\ion{O}{7} (\ion{O}{8}) line.  
The XMS full array is only slightly 
more subject to NXB and GXB uncertainties due to the poorer 
spectral resolution.

To test the non-equilibrium ionization model without ambiguity requires 
measurements of both the electron temperature (or hardness ratio) and the 
line ratio so that the measured line ratio can be compared to the CIE 
line ratio inferred by the electron temperature.  We have shown that this 
can be done within $\lesssim 2$~Mpc of a cluster by the {\it IXO} with 
sufficient confidence.  Beyond $\sim 2.5$~Mpc where the surface brightness 
may be too low and measuring the electron temperature may be difficult, we 
have shown that if the line ratio is measured to be as low as $\sim 2$ at 
3-$\sigma$ ($\Delta [S($\ion{O}{8}$)/S($\ion{O}{7}$)] \sim 2/3$ at 
1-$\sigma$) in the outermost regions, this will rule out the CIE model at 
a 3-$\sigma$ level since the CIE line ratios are always higher than 4 for 
realistic cluster temperatures.  A 3 or 4-$\sigma$ measurement of such a 
low line ratio is sufficient to provide a strong test of the 
non-equilibrium ionization.  If the line ratio is measured to be even 
lower (e.g., $\lesssim 1$), only a 2 or 3-$\sigma$ will be sufficient to 
rule out the CIE model.  On the other hand, if the line ratio is measured 
to be as high as 4, a 6-$\sigma$ measurement ($\Delta 
[S($\ion{O}{8}$)/S($\ion{O}{7}$)$] = 2/3 at 1-$\sigma$) will be necessary 
to rule out the NEI low line ratio of $\sim 2$ at a 3-$\sigma$ level.  We 
found that an observation with about 130 (220)~ksec with the XMS full 
(core) array is enough to measure the line ratio at 2.3-$\sigma$.  For a 
3-$\sigma$ measurement of the line ratio, about 230 (380)~ksec will be 
needed for the XMS full (core) array, and this will provide a strong test 
for non-equilibrium ionization.  In summary, detecting the \ion{O}{7} and 
\ion{O}{8} lines around the cluster accretion shock regions and testing 
non-equilibrium ionization in cluster outer regions with {\it IXO} are 
promising.

It is expected that the \ion{O}{7} and \ion{O}{8} lines from WHIM will 
also be strong.  Because of the high spectral resolution of the XMS, 
emissions from different redshifts should be easily separated.  Only the 
emission from WHIM immediately surrounding the target cluster will be potentially confused 
with the emission from cluster outer regions.  To observe the \ion{O}{7} 
and \ion{O}{8} lines and study NEI effects in cluster accretion shock 
regions, it will be best to avoid observing directions along the 
filaments where it is believed that denser preheated WHIMs and 
subclusters are preferentially accreted onto more massive and relaxed 
galaxy clusters. 

What do we learn about clusters from the ionization state of the outer gas?
Since collisional ionization and recombination rates involve 
straightforward atomic physics, the
processes are not in question and the rates are reasonably well-known.
Unlike shock electron heating or rates for transport processes like 
thermal conduction, the basic
physics is not uncertain and magnetic fields do not affect the results 
in a significant way.
What we mainly learn about is the pre-shock physical state of materials 
which are being
accreted by the cluster.
If most of the WHIM is ionized beyond \ion{O}{7}, then the effects 
described in this paper will be greatly reduced.
If most of the material currently being accreted by clusters comes in 
through filaments which have
a higher ionization, then NEI effects will be diminished significantly.
If most of the gas being added to clusters at present comes in through 
mergers with groups which deposit most of the gas in the inner regions 
of clusters, the gas will achieve CIE quickly.

From the theoretical point of view, with the increasing number of 
observations of galaxy cluster outer regions ($\sim R_{200}$) and the 
potential to extend observations out to the shock radius with {\it IXO} 
in the 
future, it is necessary to perform more detailed simulations than 
ours.  It is also interesting to extend our work to study the connections 
between the shocked ICM and the more diffuse WHIM surrounding clusters.  
Three-dimensional simulations will be essential to understand the effects 
of mergers or filament accretion on the degree of ionizations in 
different regions of clusters.  This will allow us to characterize the 
variation of non-equilibrium signatures in the clusters; such 
calculations are essential to compare observational signatures with our 
understanding of the cluster physics near the accretion shocks.  
Cosmological simulations have been performed recently to study the NEI 
signatures \citep{YYS+03,CF06, YS06}.  
These studies have shown that both 
non-equipartition and non-equilibrium ionization effects are important in 
cluster outer regions; although they focus more on the lower density and 
lower temperature WHIM.  High resolution simulations were also performed 
for studying NEI effects in clusters, but these are limited to binary 
mergers with idealized initial conditions and focus on the denser merger 
shocks \citep{AY08, AY10}.  Re-simulating representative clusters and the 
surrounding WHIM from cosmological simulations with higher resolutions 
and including realistic physics (e.g., cooling, conduction, turbulent 
pressure, magnetic pressure, and relativistic support by cosmic rays) 
will be necessary to provide realistic model images and spectra.  The 
different observational signatures and connections between the ICM and 
the more diffuse WHIM can also be addressed self-consistently by these 
simulations.

\acknowledgments
We thank Daniel Wang and Todd Tripp for helpful discussions.
Support for this work was provided by NASA through  {\it Chandra} grants
GO7-8129X,
GO7-8081A,
GO8-9083X,
GO9-0135X,
and
GO9-0148X,
{\it XMM-Newton} grants
NNX08AZ34G,
and
NNX08AW83G,
and
{\it Suzaku} grants
NNX08AZ99G,
NNX09AH25G,
and
NNX09AH74G.
We thank the referee for helpful comments.

\clearpage

\begin{deluxetable}{cccc}
\tabletypesize{\small}
\tablewidth{0pt}  
\tablecolumns{4}
\tablecaption{Parameters for the {\it IXO} XMS Core and Full Arrays
\label{ion_tab:IXO}
}
\tablehead{
\colhead{Instrument} &
\colhead{$A_{\rm eff}$ (cm$^2$)} &
\colhead{$\Omega_{\rm FOV}$ (arcmin$^2$)} &
\colhead{$\Delta E$ (eV)}
}
\startdata
XMSC & 10000 & $2 \times 2$ & 2.5 \\
XMSF & 10000 & $5 \times 5$ & 10
\enddata
\tablecomments{Column 3 lists the field of view.  Column 4 lists the 
spectral resolution (FWHM).
}
\end{deluxetable}

\begin{deluxetable}{cccccc}
\tabletypesize{\small}
\tablewidth{0pt}  
\tablecolumns{6}
\tablecaption{Count Rates
\label{ion_tab:bg}
}
\tablehead{
\colhead{Instrument} &
\colhead{Ion} &
\colhead{$R _{\rm line}$} &
\colhead{$R_{\rm cont}$} &
\colhead{$R _{\rm NXB}$} &
\colhead{$R_{\rm GXB}$}\\
\colhead{} &
\colhead{} &
\colhead{($10^{-4}$~counts s$^{-1}$)} &
\colhead{($10^{-4}$~counts s$^{-1}$)} &
\colhead{($10^{-4}$~counts s$^{-1}$)} &
\colhead{($10^{-4}$~counts s$^{-1}$)} 
}
\startdata
XMSC & \ion{O}{7}  & 2.11  & 0.210  & 2.44 & 6.09 \\
XMSF & \ion{O}{7}  & 11.3 & 4.03  & 46.7  & 117  \\
XMSC & \ion{O}{8}  & 1.66  & 0.171  & 0.878 & 2.19 \\
XMSF & \ion{O}{8}  & 8.69  & 1.43\   & 20.8\  & 51.8
\enddata
\end{deluxetable}

\begin{deluxetable}{ccccc}
\tabletypesize{\small}
\tablewidth{0pt}  
\tablecolumns{5}
\tablecaption{Minimum and Maximum Redshifts
\label{ion_tab:min}
}
\tablehead{
\colhead{Instrument} &
\colhead{$\Delta E_{\rm min}$} &
\colhead{$z_{\rm min}$} &
\colhead{$\Delta E_{\rm max}$} &
\colhead{$z_{\rm max}$} \\
& (eV) & & (eV) &
}
\startdata
XMSC & 15.5 & 0.0278 & 58.1 & 0.116\phn \\
XMSF & 23.0 & 0.0417 & 50.6 & 0.0992
\enddata
\tablecomments{
Column 2 lists the minimum energy shifts in order for the \ion{O}{7} 
triplets to be separated from Galactic lines.  
Column 3 lists the corresponding redshifts of $\Delta E_{\rm min}$.  
Column 4 lists the maximum energy shifts in order for the \ion{O}{7} 
triplets not to overlap with the 0.5~keV nitrogen line.
Column 5 lists the corresponding redshifts of $\Delta E_{\rm max}$. 
}
\end{deluxetable}

\clearpage

\begin{figure}
\includegraphics[angle=270,width=12cm]{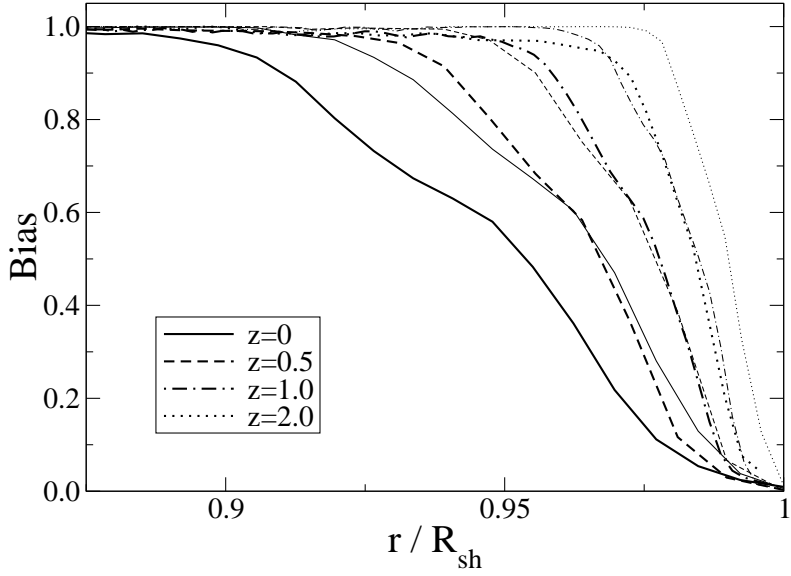}
\caption{
Non-equilibrium ionization bias versus the scaled radius ($r/R_{\rm sh}$) for
clusters at different redshifts.
The cluster model with a total accreted mass $M_{\rm sh} = 1.53 \times
10^{15}~M_{\odot}$ at $z=0$ is used.
Models with $\beta = 1/1800$ (non-equipartition) and $\beta = 1$ 
(equipartition) are shown in thick and thin lines, respectively.
}
\label{ion_fig:OfracBias_vs_z}
\end{figure}

\begin{figure}
\includegraphics[angle=270,width=12cm]{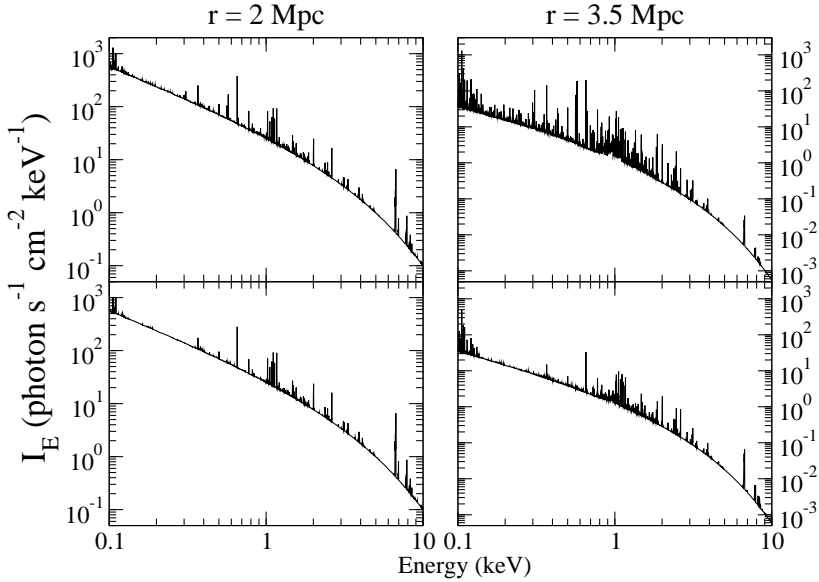}
\caption{
Projected rest frame spectra for the cluster model with an accreted mass 
of $M_{\rm sh} = 1.53 
\times 10^{15}~M_{\odot}$ at $z=0$ are shown at two projected radii (left: 
2 Mpc; right: 3.5 Mpc).
Upper panels: the NEI model with $\beta 
= 1/1800$ and non-equipartition of electrons and ions.
Lower panels: projected rest frame spectra for the 
CIE model with the same $\beta$ and non-equipartition.
}
\label{ion_fig:spec}
\end{figure}

\begin{figure}
\includegraphics[angle=270,width=12cm]{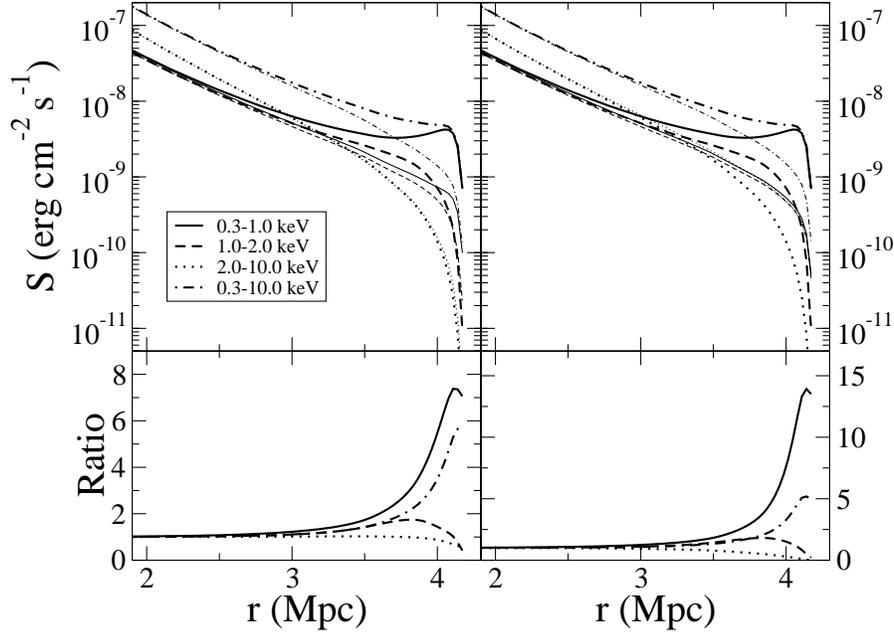}
\caption{
Upper left panel: rest frame projected surface brightness profiles for 
different energy bands for the NEI model with $\beta = 1/1800$ (thick 
lines) and the CIE--Non-Eq model (thin lines).  Upper right panel: rest 
frame projected surface brightness profiles for the NEI model with $\beta 
= 1/1800$ (thick lines) and the CIE--Eq model (thin lines).
Lower left panel: ratios of the surface brightness profiles $S_{\rm NEI}/ 
S_{\rm CIE-Non{\text -}Eq}$.  Lower right panel: ratios of the surface 
brightness profiles $S_{\rm NEI}/ S_{\rm CIE-Eq}$.
}
\label{ion_fig:sb}
\end{figure}

\begin{figure}
\includegraphics[angle=270,width=12cm]{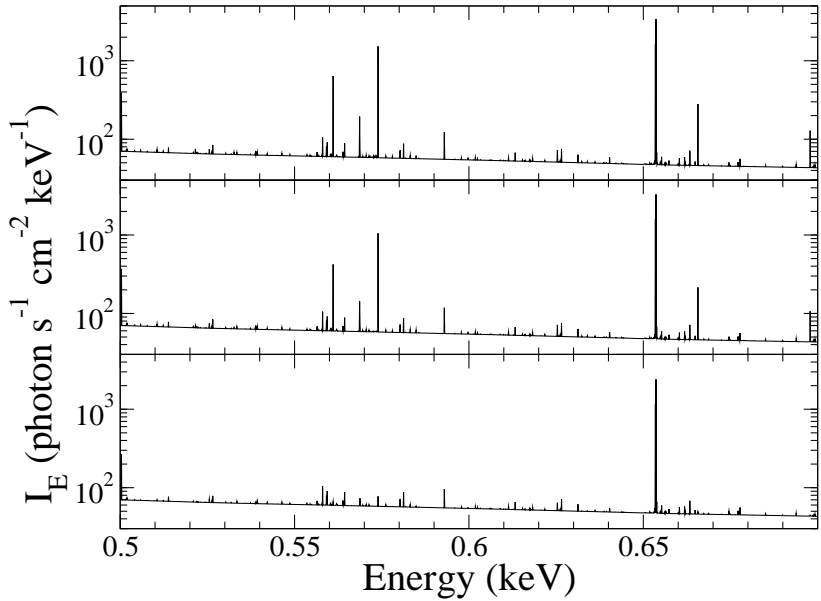}
\caption{
Upper panel: projected rest frame spectra for the NEI model with $\beta = 
1/1800$. 
Middle panel: projected rest frame spectra for the NEI model with $\beta 
= 0.5$.
Lower panel: projected rest frame spectra for the CIE--Non-Eq model with 
$\beta = 1/1800$.
All spectra are for the $M_{\rm sh} = 1.53 \times 10^{15}~M_{\odot}$ at 
$z=0$, and are for a
projected radius of $r=2$~Mpc. 
The spectra are binned with $\Delta E=0.1$~eV.
}
\label{ion_fig:spec_line}
\end{figure}

\begin{figure}
\includegraphics[angle=270,width=12cm]{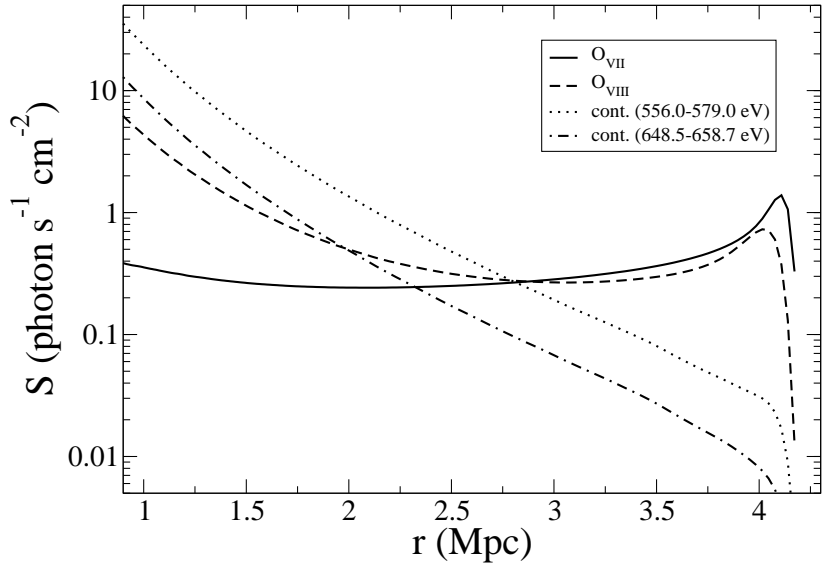}
\caption{
Surface brightness of the \ion{O}{7} and \ion{O}{8} lines for the 
NEI model 
with $\beta = 1/1800$ as a function of projected radius.
The continuum emission within the narrow energy bands covering the lines
is
also shown.
}
\label{ion_fig:line_flux}
\end{figure}

\begin{figure}
\includegraphics[angle=270,width=12cm]{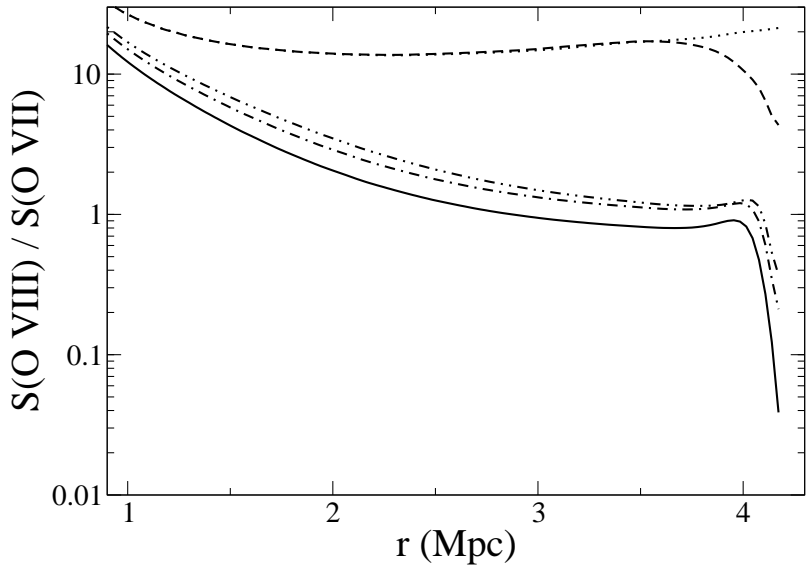}
\caption{
Line ratios $S($\ion{O}{8}$)/S($\ion{O}{7}$)$ for the NEI models with 
$\beta = 1/1800$ (solid), 0.5 (dash-dotted), and 1.0 (dash-dot-dotted).  
Line ratios for the CIE--Non-Eq (dashed) and CIE--Eq (dotted) models are 
also shown.
}
\label{ion_fig:lineratio}
\end{figure}

\begin{figure}
\includegraphics[angle=0,width=16cm]{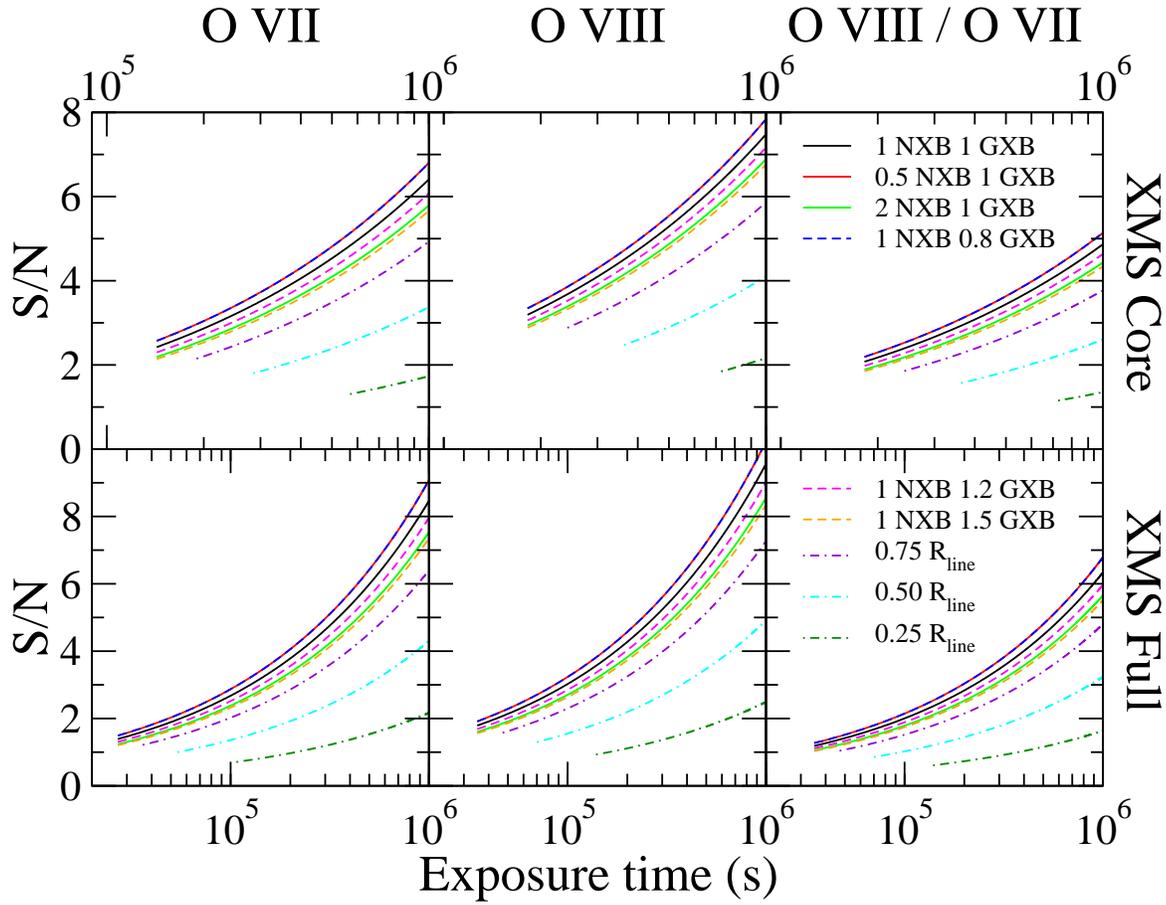}
\caption{
Left (middle) panels show the signal-to-noise ratios for the 
\ion{O}{7} triplets (\ion{O}{8} doublets) expected to be detected by 
{\it IXO}.  
The right panels show the signal-to-noise ratios for the \ion{O}{7} and 
\ion{O}{8} line ratios.
The upper panels correspond to the XMSC, and the lower
panels correspond to the XMSF.
Different signal and noise levels are assumed for the different models.  
The legends for the lines are the same for all panels, but are separated 
into two panels due to the space limitation.
}
\label{ion_fig:SN1}
\end{figure}

\end{document}